\documentclass[10pt]{article}
\pdfoutput=1
\usepackage[utf8]{inputenc}
\usepackage{cite}
\usepackage{amsmath,amssymb,amsbsy,amstext,amsthm,simplewick,amsfonts}
\usepackage{graphicx}
\usepackage{subcaption}
\usepackage{lscape}
\usepackage{cancel}
\usepackage{wrapfig}
\usepackage{upgreek}
\usepackage{bm} 
\usepackage{framed}
\usepackage{bbm}
\usepackage{textcomp}
\usepackage{tikz}
\usepackage{pifont}
\usetikzlibrary{matrix,shapes,fit,tikzmark,calc}
\usepackage{adjustbox}
\usepackage{makecell}
\usepackage{tcolorbox}
\usepackage{physics}
\usepackage{empheq}
\usepackage[normalem]{ulem}
\usepackage{enumitem}
\usepackage{braket} 
\usepackage{array}
\usepackage{dsfont}
\usepackage{ulem}
\usepackage{multirow}
\usepackage{titlesec}
\titleformat{\section}{\normalfont\fontsize{12}{16}\bfseries}{\thesection}{1em}{}

\numberwithin{equation}{section}

\def\be{\begin{equation}}
\def\ee{\end{equation}}

\def\d{\delta}

\def\e{\epsilon}

\def\ba{\begin{eqnarray}}
\def\ea{\end{eqnarray}}

\def\bfx{\textbf{x}}

\def\bfk{\textbf{k}}
\def\bfs{\textbf{s}}

\def\bfx{\textbf{x}}

\def\bfs{\textbf{s}}
\def\bfr{\textbf{r}}

\newcommand{\kmin}{k_{\text{min}}}
\newcommand{\kmax}{k_{\text{max}}}

\definecolor{blue3}{RGB}{31,119,180}
\definecolor{red3}{RGB}{214,39,40}
\definecolor{orange3}{RGB}{255,127,14}
\definecolor{green3}{RGB}{44,160,44}

\usepackage{colortbl}
\definecolor{lightgreen}{cmyk}{0.2, 0, 0.2, 0.2}
\definecolor{lightgray}{cmyk}{0.1,0.2,0,0.1}
\definecolor{lightgray2}{cmyk}{0.1,0.1,0,0.1}

\usepackage{hyperref}
\hypersetup{colorlinks=true,linkcolor=teal,citecolor=orange3,urlcolor=green3,pdfencoding=auto}

\makeatletter
\newlength{\apb@width}
\newcommand{\autoparbox}[2][c]{\settowidth{\apb@width}{#2}\parbox[#1]{\apb@width}{#2}}

\makeatother


\def\d{{\rm d}}

\def\Mpl{M_{\text{P}}}

\def\I{{\cal I}}

\def\bfp{\textbf{p}}
\def\bfs{\textbf{s}}

\def\beq{\begin{equation}}
\def\eeq{\end{equation}}
\newcommand{\ex}[1]{\langle #1 \rangle}

\newcommand{\dt}{\tilde \delta_{D}^{(3)}}

\newcommand{\ksc}{P_{\text{s.c.}}}

\newcommand{\vk}{\mathbf{k}}
\newcommand{\BPE}{B^{\text{PE}}}
\newcommand{\BPO}{B^{\text{PO}}_4}

\allowdisplaybreaks[1]
\begin{document}


\begin{titlepage}
\setcounter{page}{1} \baselineskip=15.5pt

\thispagestyle{empty}

\renewcommand*{\thefootnote}{\fnsymbol{footnote}}

\begin{center}


{\fontsize
{20}{20} \bf Leading Loops in Cosmological Correlators} \\

\end{center}

\vskip 18pt
\begin{center}
\noindent
{\fontsize{12}{18}\selectfont Mang Hei Gordon Lee\footnote{\tt  mhgl2@cam.ac.uk}, Ciaran McCulloch\footnote{\tt  cam235@cam.ac.uk}, and Enrico Pajer\footnote{\tt enrico.pajer@gmail.com}}
\end{center}

\begin{center}
\vskip 8pt
\textit{Department of Applied Mathematics and Theoretical Physics, University of Cambridge, Wilberforce Road, Cambridge, CB3 0WA, UK} 
\end{center}


\vspace{1.4cm}

\noindent Cosmological correlators from inflation are often generated at tree level and hence loop contributions are bounded to be small corrections by perturbativity. Here we discuss a scenario where this is not the case. Recently, it has been shown that for any number of scalar fields of any mass, the parity-odd trispectrum of a massless scalar must vanish in the limit of exact scale invariance due to unitarity and the choice of initial state. By carefully handling UV-divergences, 
we show that the one-loop contribution is non-vanishing and hence leading. Surprisingly, the one-loop parity-odd trispectrum is simply a rational function of kinematics, which we compute explicitly in a series of models, including single-clock inflation. Although the loop contribution is the leading term in the parity-odd sector, its signal-to-noise ratio is typically bounded from above by that of a corresponding tree-level parity-even trispectrum, unless instrumental noise and systematics for the two observables differ. Furthermore, we identify a series of loop contributions to the wavefunction that cancel exactly when computing correlators, suggesting a more general phenomenon.


\end{titlepage}

\setcounter{tocdepth}{2}
{
\hypersetup{linkcolor=black}
\tableofcontents
}

\renewcommand*{\thefootnote}{\arabic{footnote}}
\setcounter{footnote}{0} 

\newpage

\section{Introduction}
Our current leading paradigm posits that cosmological structures were seeded by quantum fluctuations of one or more scalar fields during the primordial universe. To test this hypothesis and better constrain the many proposed models, we search cosmological surveys for statistical correlations of primordial curvature perturbations. The smallness of primordial perturbations has important consequences for this search: first, the leading observable signal of interactions is expected to appear in the lowest correlation functions, such as the bispectrum (3-point) and trispectrum (4-point); second, the primordial dynamics can be described by the perturbative interactions of some weakly-coupled degree(s) of freedom; and third, the leading quantum effect consists of having a quantum state with many possible fluctuations, whose mutual interactions can be approximated as classical to leading order. In less colorful but more precise words, our theoretical predictions mostly feature tree-level contributions to low-point correlators. On general grounds of unitarity, these contributions are accompanied by loop contributions, but these are small corrections to tree-level results when the calculation is trustworthy.

In this work we investigate an interesting exception to this general expectation, namely a regime in which a primordial correlator starts at loop order. More in detail, we focus on the parity-odd sector of a scalar theory. As is well known, the scalar power spectrum and bispectrum are completely blind to any violation of parity, under the standard assumption of statistical homogeneity and isotropy. The leading probe of parity (point inversion) is the scalar four-point function, also known as \textit{trispectrum}. The parity-odd sector is interesting because it has emerged as a particularly sensitive probe of physics beyond vanilla inflationary models. Indeed, it has been noticed that the parity-odd trispectrum, which we henceforth denote by $\BPO$, vanishes in large classes of minimal models \cite{Liu:2019fag,Cabass:2022rhr}. These no-go theorems make the following assumptions about the framework of the calculation and the field content of the model. The framework assumptions are: a Bunch-Davies initial state, namely a de Sitter invariant state that reduces to the standard Lorentz-invariant Minkowski vacuum on short scales at early times; unitary time evolution; the limit of exact scale invariance; and working at tree level. The field content assumption are either (i) any number of scalar fields of any mass (always with massless or conformally coupled external scalars), or (ii) fields of any spin and massless or conformally coupled mode function with parity-even power spectra. It should be noticed that the peculiar behaviour of parity-odd correlators has avatars also in the tensor sector. In \cite{Soda:2011am} it was observed that, under the same framework assumptions as above, the parity-odd de Sitter invariant graviton cubic wavefunction coefficient computed in \cite{Maldacena:2011nz} did not contribute to correlators. For boost-breaking interactions, such as those arising in the effective field theory of inflation \cite{CabassBordin,Bartolo:2020gsh}, it was noticed that no total energy poles are generated, in contrast to general expectations. These observations were generalized in \cite{Cabass:2022rhr} and understood as simple consequences of unitary time evolution from a Bunch-Davies vacuum, via the cosmological optical theorem \cite{COT,Goodhew:2021oqg,sCOTt} (see \cite{Baumann:2022jpr} for an overview of recent related results).\\

\begin{figure}[t]
    \centering
    \begin{tabular}{c c}
       \includegraphics[scale=2]{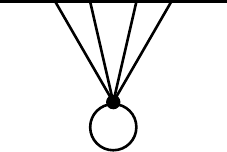} & \hspace{1cm}
       \includegraphics[scale=2]{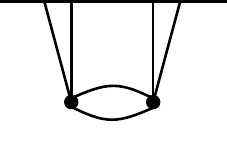}   \vspace{0.3cm} \\
       $\displaystyle{\psi_4 \sim \log k_T}$ &  \hspace{1cm} $\displaystyle{ \psi_4^{\text{PO}} = \mathbb{R} + i \frac{\text{Poly}_{p + 3} (\{\vk\})}{k_T^p} }$  \vspace{0.15cm} \\
       $\displaystyle{B_4 = 0}$ &  \hspace{1cm} $\displaystyle{B_4^{\text{PO}} \approx i \frac{\text{Poly}_{p+3} (\{\vk\})}{k_T^p \left( k_1 k_2 k_3 k_4 \right)^3}}$
    \end{tabular}
    \caption{Summary of our results. For massless or conformally-coupled scalars on de Sitter spacetime, the one-vertex one-loop wavefunction coefficient $\psi_4$  depends logarithmically on the external kinematics, but the associated one-vertex one-loop trispectrum always vanishes. On the other hand, with two vertices, one of which has a parity-odd interaction, the wavefunction coefficient $\psi_4^{\text{PO}}$ takes the form of a complicated but real-valued function (denoted $\mathbb{R}$) and a simple imaginary part, which is a rational function of the external kinematics. Only this simple imaginary part contributes to the one-loop parity-odd trispectrum.}
    \label{introfig}
\end{figure}

The goal of this paper is to investigate what happens to $\BPO$ at one loop. This requires careful handling of UV divergences. In particular, if one applied dimensional regularization (dim reg) by only extending loop integrals over momentum from $d^3p$ to $d^{d}p$, as done in early works on loop contributions \cite{Tsamis:1994ca,Weinberg:2005vy,Weinberg:2006ac}, one would find that $\BPO$ vanishes to all loops \cite{Cabass:2022rhr}. Conversely, here we adopt the prescription put forward in \cite{Senatore:2009cf}, which requires to also analytically continue the mode functions, and has the feature to maintain manifest scale invariance. In this way, we find a non-vanishing contribution to $\BPO$, which we compute explicitly in a variety of models with massless and conformally coupled scalar fields, including single-clock inflation. Our results are summarized in Figure \ref{introfig}. We find that the one-loop contribution to $\BPO$ from a diagram with a single vertex vanishes in dim reg, irrespective of parity. This is familiar from scattering amplitudes, since there is no momentum flow in the loop. Less familiar is the fact that one-loop one-vertex diagrams contribute to the wavefunction in a way that cancels out exactly when computing correlators. This is intriguing and deserves further investigation. Because of this observation, for this calculation we abandon the wavefunction and work directly with the correlators using the in-in formalism (or equivalently the Schwinger-Keldysh path integral \cite{Schwinger:1960qe,Keldysh:1964ud}). We compute the contribution from one-loop diagrams with two vertices and find a non-vanishing $\BPO$. Remarkably, for massless or conformally couple fields, the result takes the simple form
\begin{align}\label{schem}
    \BPO \sim \frac{\text{Poly}_{p+3}(\{\bfk\})}{k_T^p (k_1k_2k_3k_4)^3}\,,
\end{align}
where $k_T=\sum_a^4 k_a$, $p$ is a positive integer that depends on the number of derivatives on each of the two vertices, and the numerator is a polynomial in the external momenta. The explicit polynomial depends on the model and concrete examples can be found in \eqref{finalss} and \eqref{finalss2}, where the field in the loop is a spectator field, and \eqref{finalBPO}-\eqref{last} for single-clock inflation. Several properties of this result are surprising. First, it does not involve logarithms or polylogarithms, as one might have expected from a loop diagram, but simply a rational function. This is enormously simpler than the more general parity-even one-loop contributions \cite{Wang:2021qez,Xianyu:2022jwk,Qin:2023bjk}. 
Second, the rational function has only a pole at $k_T=0$. This is even simpler than a tree-level exchange diagram, which has also poles at vanishing partial energies. In fact, the result in \eqref{schem} looks just like that of contact diagram, albeit from a non-unitarity theory with an imaginary coupling. Third, let's stress that $\BPO$ is UV finite, even in the presence of higher-derivative operators. The UV-divergences that would be expected by power counting and that would contribute to a parity-even correlator drop out in the parity-odd counter part. This is reassuring since there is no tree-level counterterm that could re-absorb the divergence.\\

The rest of this paper is organised as follows. In Section \ref{sec:dim reg}, we review the prescription to regulate UV divergences in inflationary correlators using dimensional regularization. In Section \ref{sec:one_vertex}, we consider the simplest one-loop contribution, which arises from a diagram with a single vertex, as in Figure \ref{fig1L1V}. We show that such diagrams with loops of massless particles contribute to the wavefunction but cancel out when computing correlators (in dim reg). In Section \ref{sec:two_vertex}, we move on to the next simplest loop diagrams, namely those involving two vertices, and we focus henceforth on parity-odd contributions, for which the corresponding tree-level term vanishes in the scale invariant limit. We compute the first one-loop parity-odd trispectrum $\BPO$ for a simple toy model consisting of six distinct conformally coupled\footnote{As routinely done in the literature, by "conformally coupled" scalar we denote a scalar field whose free theory minimally coupled to gravity is invariant under a Weyl rescaling. The interactions of this scalar with itself of with other scalar fields are not invariant under this rescaling.} scalars in de Sitter. This calculation demonstrates a series of interesting and surprising features while keeping the algebra as simple as possible. Then, in Section \ref{sec:conformal} we consider $\BPO$ for a single massless scalar, which could be the inflaton, mediated by a loop of conformally coupled scalars, which could be spectators fields during inflation. Finally, we calculate $\BPO$ in single-clock inflation from self interactions in Section \ref{sec:single}. We estimate the signal-to-noise ratio for this observable in Section \ref{sec:signal} and show that it is always dominated by that of tree-level contributions to the parity-even trispectrum. Finally, in Section \ref{sec:natural} we discuss a consistent power counting scheme for scalar self-interactions in parity-violating theories. We conclude with a discussion of our results in Section \ref{sec:conclusion}.

\paragraph{Notation and conventions} We denote external momenta by $\{\bfk_1,\bfk_2,\bfk_3,\bfk_4\}$ and the loop momenta by $\bfp_1$ and $\bfp_2=-(\bfk_1+\bfk_2+\bfp_1)$. To discuss the kinematic dependence, we make use of the following compact notation
\begin{align}\label{defs}
 \omega_L &=k_1+k_2\,, & \omega_R &=k_3+k_4 \,, & \bfs &= \bfk_1+\bfk_2=-(\bfk_3+\bfk_4)\,, \\  k_T^{(n)}&=\sum_a^n k_a \,, & e_4&=k_1k_2k_3k_4\,, & p_\pm&=p_1 \pm p_2\,.
\end{align}
Momentum entering the vertex is $+\bfp$. The wavefunction is parameterezied as
\begin{align}\label{psin}
\Psi[\phi;t]=\exp\left[   +\sum_{n}^{\infty}\frac{1}{n!} \int_{\bfk_{1},\dots\bfk_{n}}\,  \dt  \left( \sum_{a}^{n} \bfk_{a} \right) \psi_{n}(\{\bfk\};t)  \phi(\bfk_{1})\dots \phi(\bfk_{n})\right]\,,
\end{align}
where
\begin{align}
\dt(\bfk)\equiv (2\pi)^{3} \delta_D^{(3)}\left(  \bfk \right)\,.
\end{align}
Most of our final results take a particularly compact form when written using the following differential operators (see e.g. \cite{Arkani-Hamed:2018kmz,Baumann:2021fxj,Hillman:2021bnk})
\begin{align}
    O^{(i)}_{k} &=1-k\partial_{\omega_i}\,,
\end{align}
where $i=L,R$ an $\omega_{L,R}$ were defined in \eqref{defs}.


\section{Dimensional regularization} \label{sec:dim reg}

In this paper, we study one-loop diagrams whose corresponding integral expressions are in general formally UV divergent. Following \cite{Senatore:2009gt}, we regulate the divergences using dimensional regularization (see also \cite{Tsamis:1994ca,Weinberg:2005vy,Weinberg:2006ac} for pioneering work on loop contributions in de Sitter). In de Sitter spacetime, dimensional regularization is not as straightforward as in flat space. Naively, we would only analytically continue the number of spatial dimensions in the momentum integral from $3$ to $d=3+\delta$. However, doing so breaks scale invariance, and this is manifest in the appearance of logarithmic terms of the form $\log(k/\mu)$ in loop diagrams, even in the absence of IR divergences. To ensure manifest scale invariance, the authors of \cite{Senatore:2009gt} suggested to analytically continue the mode functions as well. In Minkowski this would be inconsequential because the mode functions are always $e^{i\Omega t}$ with $\Omega=\sqrt{\bfk^2+m^2}$ in any number of dimensions. Conversely, in de Sitter the number of spatial dimensions appears in the index of the Hankel function, which must be carefully tracked. 

Working with Hankel functions $H_\nu$ with a general complex index $\nu(d)$ is possible but leads to complicated algebraic manipulations. To avoid this while maintaining manifest scale invariance, we will employ a trick used in \cite{sCOTt}: we analytically continue both the number of spatial dimensions \textit{and} the mass of the field in such a way that the index of the Hankel function is always $\nu=3/2$. For scalar fields, this results in the following mode functions
\begin{align}
    f_k(\eta)&=(-H\eta)^{\delta/2}\frac{H\eta}{\sqrt{2k}}e^{ik\eta} &\text{(conformally coupled scalar)},\\
    f_k(\eta)&=(-H\eta)^{\delta/2}\frac{H}{\sqrt{2k^3}}(1-ik\eta)e^{ik\eta} &\text{(massless scalar)},
\end{align}
where $\delta$ should be taken to zero at the end of the calculation. For later convenience, notice that we can write the mode functions also as
\begin{align}\label{onplane}
    f_k(\eta)&=-\frac{1}{\sqrt{2k}}(iH\partial_k)^{1+\delta/2}e^{ik\eta} &\text{(conformally coupled scalar)}\\
    f_k(\eta)&=\frac{H^2}{\sqrt{2k^3}}(iH\partial_k)^{\delta/2}(1-k\partial_k)e^{ik\eta} &\text{(massless scalar)}\,.
\end{align}
This will be useful to simplify some of the calculations.


\section{One-vertex one-loop diagrams: correlators and the wavefunction}\label{sec:one_vertex}

In this section, we discuss the simplest type of loop diagrams, namely those where the loop has a single interaction vertex and hence a single bulk-bulk propagator. This discussion is independent of the number of derivatives and does not distinguish between parity-even or parity-odd interactions. We show that the contribution of these diagrams to correlators vanishes in dimensional regularization (dim reg) in Minkowski and in de Sitter spacetime, if the fields in the loop are massless. This is somewhat analogous to what happens for amplitudes.
Conversely, one-loop one-vertex diagrams with massless fields generate non-vanishing contributions to wavefunction coefficients in general. A detailed cancellations among different terms in the wavefunction when computing correlators then ensures that these two results are compatible. As we will discuss, the physical reason is that for correlators there is no energy-momentum flow inside the loop, while for wavefunction coefficients the total energy of the diagrams flows from the boundary into the loop. 

Moreover, we generalize our analysis to massive fields running in the loop and present several explicit results. Our finding are summarised in Table \ref{tab1}. 

As a last remark, we notice that one-loop one-vertex diagrams cam be made to vanish by fiat by applying normal ordering to all interactions. While this is a possible way to bypass the calculations in this section, we find it nevertheless interesting to discuss what happens for non normal ordered interactions for at least two reasons: first this gives us a simple toy model of an exact cancellation of a term in the wavefunction when computing correlators, which could be an instance of a more general phenomenon, and because normal ordering would not remove similar contributions at higher loop order.

\begin{table}[]
    \centering
    \begin{tabular}{c|c c c|c c}
       & \multicolumn{3}{c |}{de Sitter} & \multicolumn{2}{c}{Minkowski} \\
       \hline
       & $m=0$ & $m=\sqrt{2} H$ & $m \neq 0,\, \sqrt{2}H$ & $m = 0$ & $m \neq 0$ \\
       \hline
        $\psi_4^{(1L)}$ & IR divergences? & $\log k$ & complicated  & $\log k$  & $\log m$   \\
        $B_4^{(1L)}$ & 0 & 0 & analytic & 0 & analytic
    \end{tabular}
    \caption{Summary of the results of the one-loop one-vertex diagrams. Here $\log k$ denotes schematically the logarithm of some combination of external kinematics; correlators marked `analytic' are analytic in the external kinematics; where logarithms appear during regularisation of loop integrals, they can be removed entirely by a judicious choice of the renormalisation scale. The non-analytic terms in $\psi_4$ cancel out with other non-analytic terms related to tree-level wavefunction coefficients when computing correlators.}
    \label{tab1}
\end{table}

\subsection{Correlators} Let's start computing a simple one-loop, one-vertex contribution to a correlator. For concreteness we focus on a four-point function, but the same discussion applies for any $n$-point function. For simplicity of exposition, we consider a single scalar field.

\begin{figure}[h!]
    \centering
    \includegraphics[scale=1.5]{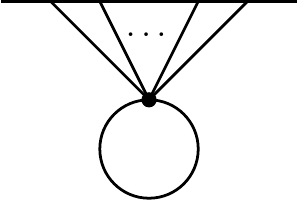}
    \caption{One-vertex one-loop diagram for a $n$-point correlator.}
    \label{fig1L1V}
\end{figure}

\paragraph{Minkowski spacetime} We start in Minkowski, and then discuss de Sitter spacetime. To use the Feynmann rules to compute correlators we need the bulk-boundary and bulk-bulk propagators, which in Minkowski are simply
\begin{align}
G_{+}(t,k)&=\frac{e^{i\Omega t}}{ 2\Omega}\,, \quad \quad G_{+-}(t_{1},t_{2},p)=G_{-+}^\ast(t_{1},t_{2},p)=\frac{e^{i\Omega (t_{1}-t_{2})}}{2\Omega}\,,\\
G_{++}(t_{1},t_{2},p)&=\frac{e^{i\Omega (t_{2}-t_{1})}}{2\Omega}\theta(t_{1}-t_{2})+\left( t_{1}\leftrightarrow t_{2} \right)\,.
\end{align}
where $\Omega=\sqrt{k^2+m^2}$ is the on-shell energy and the labels $\pm$ refer to interactions in the time ordered time evolution of the ket or anti-time ordered time evolution of the bra in the in-in correlator. Since our results will not depend on the number of spatial or time derivatives, we consider a simple polynomial interaction,
\begin{align}\label{phi6}
    \mathcal{L}_{\text{int}}=\int_\bfx \, \frac{\lambda}{6!}\phi^{6}\,.
\end{align}

For the diagram in Figure \ref{fig1L1V} we have 
\begin{align}
   B_4&=2\Re \left[ \frac{i\lambda}{2} \int_\bfp \int_\infty^0 dt \, G_{++}(t,t,p)\prod_a^4 G_+(t,k_a)\, \right] \,.
\end{align}
The crucial point is that, since the bulk-bulk propagator is evaluated at coincident times, the oscillating exponentials cancel each other and the dependence on loop momentum is only given by the overall normalization factor
\begin{align}
G_{++}(t,t,p)&=\frac{e^{i\Omega (t-t)}}{2\Omega}=\frac{1}{2\sqrt{p^2+M^2}} \overset{M \to 0}{\longrightarrow} \frac{1}{2p}\,.
\end{align}
For a massless scalar this reduces to a power law dependence, $G_{++}\propto 1/p$. Notice that the dependence would still be a power law in the presence of time and space derivatives from local interactions. Now we regulate the loop integral in $\bfp$ using dim reg. Since the integral is a power of $p$, it vanishes in dim reg\footnote{Had we used a cutoff regularization we would have found power law divergences, to be removed by renormalization, but no left over logarithmic running}
\begin{align}
\int dp^d \, p^\alpha =0 \quad \text{(dim reg)}\,.
\end{align}
This is intuitive because there is no scale in the integrand with which to write a dimensionally correct result. We conclude that a loop of a massless particle with a single vertex does not contribute to Minkowski correlators. This would remain true if we computed the correlator at unequal times. 

If the field running in the loop is massive the momentum integral no longer vanishes in dim reg. Instead we have the following:
\begin{equation}
    B_4=\frac{1}{8 \Omega_1\Omega_2\Omega_3\Omega_4 }\int_{\bfp}\frac{\lambda}{4\Omega_p \Omega^{(4)}_T},
\end{equation}
where $\Omega_p=\sqrt{p^2+m^2}$ and $\Omega_T^{(4)} = \Omega_1 + \Omega_2 + \Omega_3 + \Omega_4$ is the total energy entering the diagram. This integral can be evaluated to give:
\begin{equation}\label{B4massive}
    B_4=\frac{1}{8 \Omega_1\Omega_2\Omega_3\Omega_4}\frac{\lambda m^2}{16\pi^2k\Omega_T^{(4)}}\left(\frac{1}{\delta}+\log\frac{m}{\mu}+\text{(analytic)}\right),
\end{equation}
where $\mu$ is a renormalization scale. Notice that the result is analytic in the external kinematics. 


\paragraph{Massless scalars on de Sitter spacetime} Something very similar happens for massless fields in de Sitter spacetime. The in-in correlator is given by:
\begin{equation}
   B_4=2\Re \left[\frac{i\lambda}{2}\int_{\bfp}\int_{-\infty}^{0}\frac{d\eta}{\eta^{d+1}}G_{++}(\eta,\eta,p)\prod_a^4 G_{+}(\eta,k_a)\right]\,.
\end{equation}
At coincident times, the bulk-to-bulk propagator reads:
\begin{equation}
    G_{++}(\eta,\eta,p)=\frac{H^2}{2p^3}(1-ip\eta)(1+ip\eta)e^{ip(\eta-\eta)}=\frac{H^2}{2p^3}(1+p^2\eta^2).
\end{equation}
The propagator is simply a polynomial in $p$, so \textit{in dim reg this vanishes} just like the Minkowski correlator. A similar cancellation also occurs for conformally coupled scalars in de Sitter as well. The vanishing of this contribution is familiar from scattering amplitudes and is usually described by saying that there is no flow of energy or momentum through the loop from the external kinematics. In the absence of both a mass and external kinematics, the loop has no way to satisfy dimensional analysis and must hence vanish. 

\paragraph{Massive scalars on de Sitter spacetime}
Similarly to the case of massive scalars on Minkowski spacetime, the one-loop one-vertex diagram on de Sitter is not expected to vanish for massive fields.
The mode function for a massive scalar on de Sitter in the dim reg procedure described in Section \ref{sec:dim reg} is
\begin{equation}
   f_k(\eta) = \frac{i \sqrt{\pi} H^{1 + \frac{\delta}{2}}}{2} (-\eta)^{\frac{3 + \delta}{2}} H_{\nu} ^{(1)}(-k\eta),
\end{equation}
with $\nu = \sqrt{\frac{9}{4} - \frac{m^2}{H^2}}$ and $H^{(1)}$ the Hankel function of the first kind. The one-loop trispectrum for a massive scalar with a $\lambda\phi^6/6!$ interaction at conformal time $\eta_f$ is then given by the following integral:
\begin{multline}
   B_4 = \Re \left(  i \lambda \frac{ \pi^4  H^{8 + 4\delta}}{256} (-\eta_f)^{2(3 + \delta)} H_{\nu} ^{(1)\, *}(-k_1\eta_f) \dots H_{\nu} ^{(1)\, *}(-k_4\eta_f) \right.\\
   \left. \int_{\vb p} \int_{-\infty}^{\eta_f} \dd \eta a^{4 + \delta}(\eta) (-\eta)^{2(3 + \delta)} H_{\nu} ^{(1)}(-k_1\eta) \dots H_{\nu} ^{(1)}(-k_4\eta)  \frac{\pi H^{2 + \delta}}{4} (-\eta)^{3 + \delta} \abs{ H_{\nu} ^{(1)}(-p\eta) }^2 \right).
\end{multline}
As $H_\nu^{(1)}(x) \sim x^{-\nu}$ as $x \rightarrow 0$, we must rescale the correlator in order to find a finite value as $\eta_f \rightarrow 0$.
The resulting correlator is
\begin{multline}
   B_4' = \Re \left(  i \lambda \frac{ \pi^4  H^{8 + 4\delta}}{256} \left( \frac{2^\nu \Gamma(\nu)}{\pi} \right)^4 \right.\\
   \left. \int_{\vb p} \int_{-\infty}^{0} \dd \eta a^{4 + \delta}(\eta) (-\eta)^{2(3 + \delta)} H_{\nu} ^{(1)}(-k_1\eta) \dots H_{\nu} ^{(1)}(-k_4\eta)  \frac{\pi H^{2 + \delta}}{4} (-\eta)^{3 + \delta} \abs{ H_{\nu} ^{(1)}(-p\eta) }^2 \right).
\end{multline}

Since the momentum integral vanishes in dim reg for massless and conformally-coupled scalars, and because it contains fewer Hankel functions, it is reasonable to attempt that integral first:
\begin{equation}
   \mathcal{I}_p = \int \frac{\dd^{3 + \delta} p}{\left( 2\pi \right)^3} \abs{ H_{\nu} ^{(1)}(-p\eta) }^2.
\end{equation}
For $\nu=\frac{3}{2}$ (massless) and $\nu=\frac{1}{2}$ (conformally-coupled), the integrand is a sum of power laws in $p$ and vanishes in dim reg.
Strictly, however, $\mathcal{I}_p$ converges only for $\Re(\delta) < -2$ and $-3 -\delta < \Re(2\nu) < 3 + \delta$ (due to the behavior of the Hankel function as $p\rightarrow \infty$ and $p \rightarrow 0$ respectively).
Then,
\begin{multline}
    \mathcal{I}_p = \frac{2^{2+\delta}\csc(\pi\nu)\sec(\frac{\pi\delta}{2})}{(2\pi)^{3 + \delta} \left( - \eta \right)^{3 + \delta}\Gamma(-\frac{1}{2}-\frac{\delta}{2})}\left[-\Gamma\left(\frac{3+\delta}{2}-\nu\right)\frac{\,_2F_1\left(\frac{3+\delta}{2},\frac{3+\delta}{2}-\nu,1-\nu;1\right)}{\Gamma(1-\nu)}\sin(\frac{\pi\delta}{2}-\pi\nu)\right.\\+\left.\Gamma\left(\frac{3+\delta}{2}+\nu\right)\frac{\,_2F_1\left(\frac{3+\delta}{2},\frac{3+\delta}{2}+\nu,1+\nu;1\right)}{\Gamma(1+\nu)}\sin(\frac{\pi\delta}{2}+\pi\nu)\right].
\end{multline}
When $\Re(\delta)<-2$ this expression can be simplified as:
\begin{equation}
     \mathcal{I}_p = \frac{2^{2+\delta}\pi\csc(\pi\nu)\sec(\frac{\pi\delta}{2})\Gamma(-2-\delta)}{(2\pi)^{3 + \delta} \left( - \eta \right)^{3 + \delta}\Gamma(-\frac{1}{2}-\frac{\delta}{2})\Gamma(-\frac{1}{2}-\frac{\delta}{2}-\nu)\Gamma(-\frac{1}{2}-\frac{\delta}{2}+\nu)}\left[\frac{\sin(\frac{\pi\delta}{2}-\pi\nu)}{\cos(\frac{\pi\delta}{2}-\pi\nu)}-\frac{\sin(\frac{\pi\delta}{2}+\pi\nu)}{\cos(\frac{\pi\delta}{2}+\pi\nu)}\right].
\end{equation}
We analytically continue this expression in $\delta$ and study its behavior as $\delta\rightarrow 0$. One can check that this expression indeed gives zero for a massless or a conformally-coupled scalar, as expected. More generally, the $\Gamma(-2-\delta)$ term in the integral will contribute a $\delta^{-1}$ divergence in dim reg.


From the factors of $H^\delta$ in the time integral and the $\delta^{-1}$ divergence in the momentum integral, terms like $\log\frac{H}{\mu}$ will appear in the final correlator.
However, as long as the time integral is IR-convergent, i.e. $\Re \nu < \frac{\delta + 3}{4}$, scale invariance is unbroken, which fixes the form of the correlator and precludes any $\log \frac{k}{\mu}$ terms, where $k$ stands for some combination of the external momenta.
As on Minkowski spacetime, the resulting trispectrum must then be analytic in the external kinematics.


\subsection{Wavefunction coefficients} Now let's try and perform the same calculation using the wavefunction formalism. 

\paragraph{Massless scalar in Minkowski spacetime} Let's start with massless scalars.  In flat spacetime the wavefunction propagators are
\begin{align}\label{Gbb}
   K_{\vb k}(t) &= e^{ikt}\,, & 
   G_{\vb k}(t_1, t_2) &= \frac{1}{2ik}\left( e^{ik(t_2 + t_1)} - e^{ik(t_2 - t_1)} \right) \theta(t_1 - t_2) + (t_1 \leftrightarrow t_2).
\end{align}
With an interaction $\lambda \phi^6/6!$, the relevant wavefunction coefficients are\footnote{Notice that given our definition of the bulk-bulk propagator in \eqref{Gbb}, which includes a factor of $i$, the correct Feynman rule is to introduce a factor of $i^{1-L}$, where $L$ is the number of loops, and no factor of $i$ on the vertices, which simply result in a factor of $\lambda$.}
\begin{align}
   {\psi_6^{\text{tree}}(\vb k_1, \dots, \vb k_6)} &= i \int_{-\infty(1 - i\varepsilon)}^{0} \dd t \, e^{i k_T^{(6)} t} \lambda\\
   \label{eq:psi6T}
   &= \frac{\lambda}{k_T^{(6)}}, \quad \text{and}\\
   {\psi_4^{(1L)}(\vb k_1, \dots, \vb k_4)} &= \int^{0} \dd t \, \frac{\lambda}{2} e^{i k^{(4)}_T t} \int_{\vb p} \frac{1}{2ip}\left[ e^{2ipt} - 1 \right]. \quad \text{Performing the time integral first,}\\ \label{psi4}
   &= \int_{\vb p} \frac{\lambda}{-4p}\left[ \frac{1}{k_T^{(4)} + 2p} - \frac{1}{k_T^{(4)}}  \right].
\end{align}
This last integral is UV divergent and needs to be regularized. In dimensional regularisation, the second term in brackets in ${\psi_4^{(1L)}(\vb k_1, \dots, \vb k_4)}$ vanishes; while the first term in $\overline{MS}$, gives
\begin{equation}
   \label{eq:psi4L}
   {\psi_4^{(1L)}(\vb k_1 \dots \vb k_4)} = -\frac{6\lambda}{32\pi^2}k_T^{(4)} \ln\frac{k_T^{(4)}}{\mu}.
\end{equation}

This non-vanishing result is intriguing because we had just found that a similar 1-loop 1-vertex contribution vanishes for correlators. We will show shortly show that the two results are compatible and that indeed the term in \eqref{eq:psi4L} cancels exactly with another term when computing $B_4$. Here we would like to make some general remarks. 
Notice that in the wavefunction calculation there is a \textit{flow of energy from the external kinematics though the loop}. This is visible in the denominator $k_T^{(4)}+2p$ in \eqref{psi4} arising after performing the time integrals. This is naively surprising because we are computing a diagram that is identical to that for the correlator where we stated that there is no energy-momentum flow through the loop. The resolution is that the wavefunction, in contrast to a correlator, provides the answer to a boundary value problem where $\phi$ has been specified at some time, which we take to be $t=0$ here. This explicit boundary condition breaks time translation invariance and energy can flow from this boundary. Indeed, it is precisely the total energy that flows into the loop, because the boundary is attached to all external legs. Also, since the boundary does not break spatial translations, there is no flow of spatial momentum through the loop, only energy. At the mathematical level, the origin of the energy flow through the loop is the boundary term in the wavefunction's bulk-bulk propagator $G$, which is absent in the correlator's bulk-bulk propagator $G_{++}$. A more colorful way to say this is that the bulk-bulk propagator in the loop represents the quantum fluctuation of a virtual particle. In the correlator, such fluctuations are unconstrained, but in the wavefunction they must obey the boundary condition that $\phi$ takes some fixed value at $t=0$. This requires the quantum fluctuation to turn off as the interaction vertex is pushed toward $t=0$, which in turn requires knowledge of this fix boundary and hence a breaking of time translations. This mechanism is actually closerly related to how the recursion relations for the Minkowksi wavefunction were derived in \cite{Arkani-Hamed:2017fdk}.\\

Now use the wavefunction coefficients to find the trispectrum. Performing the average over $\phi$ in the Born rule we find
\begin{align}\label{B4}
B_{4}=\frac{1}{\prod_{a}^{4}2\Re \psi_{2}(k_{a})} \left[ \rho^{(1L)}(\{\bfk\}) + \int_{\bfp}\frac{\rho^\text{tree}(\{\bfk\},\bfp,-\bfp) }{2\Re\psi_{2}(p)}\right]
\end{align}
where $\{\bfk\}=\{\bfk_1,\dots,\bfk_4\}$ and $\rho$ denotes the coefficient of the diagonal part of the density matrix $|\Psi|^2$,
\begin{align}
   \rho^{(1L)}(\{\bfk\})&=\psi_{4}^{(1L)}(\bfk_1,\dots,\bfk_4)+\psi_{4}^{(1L)}(-\bfk_1,\dots,-\bfk_4)^{\ast} \\
   \rho^\text{tree}(\{\bfk\},\bfp,-\bfp)&=\psi_{6}^\text{tree}(\bfk_1,\dots,\bfk_4,\bfp,-\bfp)+\psi_{6}^\text{tree}(-\bfk_1,\dots,-\bfk_4,-\bfp,\bfp)^{\ast} \,.
\end{align}
The free power spectrum in Minkowski is $1/2k$ and so $\Re \psi_2=- k$. For the parity even contribution in \eqref{phi6} we can simply drop the minus sign on the momenta. Then, the first contribution to $B_4$ in \eqref{B4} is
\begin{align}
   -\frac{1}{\prod_{a}^{4}2\Re \psi_{2}(k_{a})}   \rho^{(1L)}(\{\bfk\}) &= \frac{1}{16 e_4} 2\Re \psi_4^{(1L)} (\vb k_1, \dots, \vb k_4) \\
   &= \frac{1}{8 e_4} \cdot \frac{- 6 \lambda}{32 \pi^2} k_T^{(4)} \ln\frac{k_T^{(4)}}{\mu}.
\end{align}
The second is
\begin{align}
   -\frac{1}{\prod_{a}^{4}2\Re \psi_{2}(k_{a})}   \int_{\bfp}\frac{\rho^\text{tree}(\{\bfk\},\bfp,-\bfp) }{2\Re\psi_{2}(p)} &= \frac{1}{16 e_4} \frac{1}{2} \int_{\vb p} \frac{1}{2p} 2\Re \psi_6^{\text{tree}} (\vb k_1, \dots, \vb k_4, \vb p, \vb p) \\
   &=\frac{1}{8 e_4} \frac{1}{2} \int_{\vb p} \frac{1}{2p} \frac{\lambda}{k_T^{(4)} + 2p}.
\end{align}
The momentum integral is just $-1$ times that of ${\psi_4^{(1L)} (\vb k_1 \dots \vb k_4})$, so the two contributions to the trispectrum cancel. This cancellation is interesting and deserves further investigation. 

\paragraph{Massive scalar in Minkowski spacetime} For massive scalars we no longer expect the contribution from $\rho^{(1L)}$ and $\rho^{\text{tree}}$ to cancel. Let us calculate $\psi_{\bfk_1\dots\bfk_4}^{(1L)}$ explicitly. We have:
\begin{align}
    \psi_{\bfk_1\dots\bfk_4}^{(1L)}&=\int^{0}dt\frac{\lambda}{2}e^{i \Omega_T^{(4)}t}\int_{\bfp}\frac{1}{2i\Omega_p}[e^{2i\Omega_p t}-1]\nonumber\\
    &=\frac{\lambda}{2\Omega_T^{(4)}}\int_{\bfp}\frac{1}{\Omega_T^{(4)}+2\sqrt{p^2+m^2}}.
\end{align}
In the regime $m>\Omega_T^{(4)}$ this integral can evaluated easily, since we can write the integral as:
\begin{equation}
    \psi_{\bfk_1\dots\bfk_4}^{(1L)}=\frac{\lambda}{8\pi^2\Omega_T^{(4)}}\int_{0}^{\infty}dp\frac{p^{2+\delta}}{\sqrt{p^2+m^2}}\sum_{n=0}^{\infty}\left(\frac{-\Omega_T^{(4)}}{2\sqrt{p^2+m^2}}\right)^n.
\end{equation}
Evaluating this integral gives
\begin{align}
    \psi_{\bfk_1\dots\bfk_4}^{(1L)}&=
    \sum_{n=0}^{\infty}\frac{\lambda}{16\pi^2\Omega_T^{(4)}}\left(\frac{-\Omega_T^{(4)}}{2}\right)^n m^{2-n-\delta}\frac{\Gamma(\frac{3}{2})\Gamma(\frac{n}{2}+\delta-1)}{\Gamma(\frac{n+1}{2})}\nonumber\\
    &=\frac{\lambda}{16\pi^2\Omega_T^{(4)}}\left(m^2-\frac{(\Omega^{(4)}_T)^2}{2}\right)\left(\frac{1}{\delta}+\log\frac{m}{\mu}+\text{(analytic)}\right).
\end{align}
Compared to $B_4$ we have an extra contribution of the form $\Omega_T^{(4)}\log m$, and we expect this to be cancelled by the term from $\rho^{\text{tree}}$. Indeed, we find that for $m>\Omega_T^{(4)}$:
\begin{align}
    \frac{1}{\prod_{a}^{4}2\Re \psi_{2}(k_{a})}\int_{\bfp}\frac{\rho^{\text{tree}}(\{\bfk\},\bfp,-\bfp) }{2\Re\psi_{2}(p)} &=\frac{1}{16e_4}\frac{1}{2}\int_{\bfp}\frac{1}{2\Omega_p}2\Re \psi_{6\bfk_1\dots\bfk_4\bfp\bfp}^{\text{tree}}\nonumber\\
    &=\frac{1}{8e_4}\frac{\lambda}{2}\int_{\bfp}\frac{1}{2\sqrt{p^2+m^2}}\frac{1}{\Omega_T^{(4)}+2\sqrt{p^2+m^2}}\nonumber\\
    &=\frac{1}{8e_4}\sum_{n=0}^{\infty}\frac{\lambda}{32\pi^2}\left(\frac{-\Omega_T^{(4)}}{2}\right)^n m^{1-n+\delta}\frac{\Gamma(\frac{3}{2})\Gamma(\frac{n-1}{2}-\delta)}{\Gamma(\frac{n}{2}+1)}\nonumber\\
    &=\frac{1}{8e_4}\frac{\lambda}{16\pi^2}\frac{\Omega_T^{(4)}}{2}\left(\frac{1}{\delta}+\log\frac{m}{\mu}+\text{(analytic)}\right).
\end{align}
Therefore, using \eqref{B4}, the contributions of the form $\Omega_T^{(4)}\log m$ cancels in $B_4$, and we obtain the expression in \eqref{B4massive}. It would be nice to have a systematic understanding of these type of cancellations.


\paragraph{De Sitter spacetime} We can consider a similar $\lambda \sigma^6 / 6!$ interaction of a conformally-coupled scalar on de Sitter. Since such a field is massive, as $\eta_0 \to 0$ it decays.
Formally, to avoid this issue, we consider the wavefunction of the re-scaled field $\sigma/\eta_0$ in this limit; this amounts to factoring out all factors of $\eta_0$ in the propagators. With this prescription, and using the scale-invariant dim reg procedure discussed in Section \ref{sec:dim reg}, the wavefunction propagators are
\begin{align}
   K_{\vb k}(\eta) &= (-\eta)^\delta \eta e^{ik\eta}\\
   G_{\vb k}(\eta_1, \eta_2) &= i \frac{\left( H^2 \eta_1 \eta_2 \right)^{1 + \frac{\delta}{2}}} {2k} \left[ e^{-ik(\eta_2 - \eta_1)} \theta(\eta_1 - \eta_2) + (\eta_1 \leftrightarrow \eta_2) - e^{ik(\eta_1 + \eta_2)} \right].
\end{align}
The relevant wavefunction coefficients are
\begin{align}
   {\psi_6^{\text{tree}}(\vb k_1, \dots, \vb k_6)} &= i \lambda \int_{-\infty(1 - i\varepsilon}^{0} \dd \eta \frac{1}{(-H \eta)^{4 + \delta}} (-\eta)^{6 + 3\delta} e^{i k_T^{(6)} \eta}\\
      &= -\frac{\lambda e^{2 \pi i \delta} \Gamma(3 + 2\delta)}{ H^{4 + \delta} \left( k_T^{(6)} \right)^{3 + 2\delta} }
\end{align}
and
\begin{align}
   {\psi_4^{(1L)}(\vb k_1 \dots \vb k_4)} &= \frac{\lambda}{2} \int_{-\infty(1 - i\varepsilon)}^{0} \dd \eta \frac{1}{(-H \eta)^{4 + \delta}} (-\eta)^{4 + 2\delta} e^{i k_T^{(4)} \eta}  \int \frac{\dd^{3 + \delta} p}{(2\pi)^{3+\delta}} \frac{i \left( H^2 \eta^2 \right)^{1 + \frac{\delta}{2}}}{2p} \left[ 1 - e^{2 i p \eta} \right].
\end{align}
The first term in the brackets yields a power law in $p$ after the time integral is performed, and vanishes in dim reg.
The time integral in the second term is essentially the same as in $\psi_6^{\text{tree}}$:
\begin{align}
   {\psi_4^{(1L)}(\vb k_1 \dots \vb k_4)} &= \frac{\lambda}{2} \int \frac{\dd^{3 + \delta} p}{(2\pi)^{3 + \delta}} \frac{H^{2(1 + \frac{d}{2})}}{2p} \frac{\Gamma(3 + 2\delta) e^{2 \pi i \delta}} {H^{4 + \delta} \left(k_T^{(4)} + 2p \right)^{3 + 2\delta} }.
\end{align}
This momentum integral is finite:
\begin{align}
   {\psi_4^{(1L)}(\vb k_1 \dots \vb k_4)} &= \frac{\lambda e^{2 \pi i \delta}}{2^{8 + 3\delta} \pi^{\frac{3 + \delta}{2}}  H^2 \left( k_T^{(4)} \right)^{1 + \delta}} \frac{ \Gamma(2 + \delta) \Gamma(1 + \delta)}{\Gamma(\frac{3 + \delta}{2})};
\end{align}
since it is analytic in the momenta as $\delta \rightarrow 0$, it could be removed by local counter-terms.
Local counter-terms would then also have to be added to remove the contribution of $\psi_6^{\text{tree}}$ to the trispectrum.

More generally, consider either conformally coupled scalars or massless scalar with IR finite interactions, i.e. the resulting correlators do not diverge as $\eta_0\to 0$. The integrals encountered when considering the one-vertex one-loop diagram for these fields can be grouped into two types: the first type is
\begin{equation}
    \mathcal{I}^1_m=\int_{-\infty}^{0}d\eta \,\,\eta^{n+m+2\delta}e^{ik_T^{(4)}\eta}\int_{0}^{\infty}dp\frac{p^{2+m+\delta}}{2p^3},
\end{equation}
which vanishes in dim reg. The second type is:
\begin{align}
    \mathcal{I}^2_m&=\int_{-\infty}^{0}d\eta \,\,\eta^{n+m+2\delta}e^{ik_T^{(4)}\eta}\int_{0}^{\infty}dp \frac{p^{2+m+\delta}}{2p^3}e^{2ip\eta}\nonumber\\
    &=\int_{0}^{\infty}dp\frac{(-i)^{n+m+\delta}\Gamma(n+m+1+2\delta)p^{-1+m+\delta}}{(k_T^{(4)}+2p)^{n+m+1+2\delta}},
\end{align}
where $m=0,1,2$. By power counting this momentum integral is always convergent for $n\geq 0$, which is always true for IR finite interactions. Hence terms coming from these integrals are always finite and analytic in $k_T^{(4)}$.

When we consider more general massive scalars, the integrals involved are much harder to solve. Namely, we have to integrate over products of Hankel functions, and we expect the result not to be analytic in $k_T^{(4)}$ in general. 


\section{Two-vertex one-loop diagram: general strategy and a toy model}\label{sec:two_vertex}

\begin{figure}[t]
    \centering
    \includegraphics[scale=2.5]{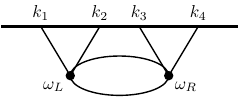}
    \caption{Two-vertex one-loop diagram for the trispectrum.}
    \label{fig:2siteloop}
\end{figure}

From now on and for the rest of the paper we will be focusing on the parity-odd trispectrum generated by the one-loop two-vertex diagram in Figure \ref{fig:2siteloop}. We will first discuss it in general and then present a series of explicit calculation in increasing order of complexity, culminating with the case of single-clock inflation.\\

To begin, let's derive an integral expression for the diagram in Figure \ref{fig:2siteloop}. To this end, consider a general interaction Hamiltonian of the form
\begin{equation}
    H_{\text{int}}(\{\bfk\},\eta)=\int_{-\infty}^{0}d\eta \left[ F_{\text{PO}}(\{\bfk\},\eta)+F_{\text{PE}}(\{\bfk\},\eta) \right] \phi(k_a)\phi(k_b)\phi(k_c)\phi(k_d),
\end{equation}
where $F_{\text{PO},\text{PE}}$ denote the vertices corresponding to a local interaction with an odd or even number of spatial derivatives, respectively, of which examples will be given later on. We can use the Feynman rules outlined in Appendix \ref{app:A} to write the trispectrum as: 
\begin{multline}\label{odd}
   B_{4}(k_1,k_2,k_3,k_4)=\sum_{a,b=\pm}\int_{-\infty}^{\eta_0}d\eta_1\int_{-\infty}^{\eta_0}d\eta_2\int_{\bfp}\delta^{(3)}(\bfp_1+\bfp_2+\bfs) G_{a}(k_1,\eta_1)G_{a}(k_2,\eta_1)\\\times F_{\text{PO}}(\bfk_1,\bfk_2,\eta_1) G_{ab}(p_1,\eta_1,\eta_2)G_{ab}(p_2,\eta_1,\eta_2)F_{\text{PE}}(\bfk_3,\bfk_4,\eta_2)G_{b}(k_3,\eta_2)G_{b}(k_4,\eta_2)\,,
\end{multline}
where we used 
\begin{align}
\bfs=\bfk_1+\bfk_2=-\bfk_3-\bfk_4   \,.
\end{align}

In-in diagrams are related pairwise. If $D$ represents a diagram with a particular choice of vertices on the $+$ or $-$ contours (from the time evolution of the bra and the ket), and $\bar D$ represents a diagram when each vertex sit on the opposite contour, $+ \leftrightarrow -$, then 
\begin{align}\label{DDbar}
    D=\bar D (-1)^n\,,
\end{align}
with $n$ the number of spatial derivatives. This ensures that in Fourier space parity-even correlators are real and parity-odd correlators are purely imaginary, as it should be for Hermitian operators in position space. Since we are considering a contribution with an overall odd number of spatial derivatives, we only need the imaginary part of the integral in \eqref{odd}. We can then write
\begin{equation}
    B_{4}=B_{4A}+B_{4B},
\end{equation}
where, for future convenience, we separate the trispectrum into two contributions,
\begin{multline}
   B_{4A}=2\text{Im}\int_{-\infty}^{\eta_0}d\eta_1\int_{-\infty}^{\eta_0}d\eta_2\int_{\bfp}\delta^{(3)}(\bfp_1+\bfp_2+\bfs) G_{+}(k_1,\eta_1)G_{+}(k_2,\eta_1)\\\times F_{\text{PO}}(\bfk_1,\bfk_2,\eta_1) G_{++}(p_1,\eta_1,\eta_2)G_{++}(p_2,\eta_1,\eta_2)F_{\text{PE}}(\bfk_3,\bfk_4,\eta_2)G_{+}(k_3,\eta_2)G_{+}(k_4,\eta_2)\,,
\end{multline}
\begin{multline}
   B_{4B}=2\text{Im}\int_{-\infty}^{\eta_0}d\eta_1\int_{-\infty}^{\eta_0}d\eta_2\int_{\bfp}\delta^{(3)}(\bfp_1+\bfp_2+\bfs) G_{+}(k_1,\eta_1)G_{+}(k_2,\eta_1)\\\times F_{\text{PO}}(\bfk_1,\bfk_2,\eta_1) G_{+-}(p_1,\eta_1,\eta_2)G_{+-}(p_2,\eta_1,\eta_2)F_{\text{PE}}(\bfk_3,\bfk_4,\eta_2)G_{-}(k_3,\eta_2)G_{-}(k_4,\eta_2).
\end{multline}


There is one last general result that will be very useful in the following explicit calculations. We will often encounter integrals of the following form:
\begin{equation}
    \int_{-\infty}^{\eta_0}d\eta \,(-H\eta)^{n+\delta}e^{ik\eta}=(iH\partial_k)^{n+\delta} \int_{-\infty}^{\eta_0}d\eta\, e^{ik\eta},
\end{equation}
where $n$ is an integer. This tells us that the trispectrum can often be written in terms of derivative operators acting on a simpler integral. In dimensional regularization, this leads to the following simplification. Suppose we want to evaluate
\begin{equation*}
    2\text{Im}\,(\partial_k)^{n}(iH\partial_k)^{\delta} I(k),
\end{equation*}
and $I(k)$ is the result of a UV-divergent integral, which can be written as:
\begin{equation*}
    I(k)=\frac{I_0(k)}{\delta}+I_1(k)+\mathcal{O}(\delta).
\end{equation*}
For the cases we will interested in, where IR divergences are absent, $I(k)$, $I_0(k)$ and $I_1(k)$ are all real as consequence of unitarity \cite{Cabass:2022rhr}. Then we can expand the derivative operator in the following way:
\begin{equation*}
    (iH\partial_k)^{\delta}=1+\delta\log(iH\partial_k)+\dots=1+\delta\left(\log(H\partial_k)+\frac{i\pi}{2}\right)+\dots.
\end{equation*}
Here the logarithm is understood as a power series in $\partial_k$. The terms from $\log(H\partial_k)$ acting on $I(k)$ are all real, so if we want to isolate the imaginary part, we find that:
\begin{equation}\label{ipidelta}
    2\text{Im}\,(\partial_k)^{n}(iH\partial_k)^{\delta} I(k)=\pi (\partial_k)^{n}I_0(k).
\end{equation}
In other words, only the coefficient of the $1/\delta$ part of the simpler integral $I(k)$ contributes to the final result. Since we only want the imaginary part when we compute the parity-odd trispectrum, we will only need to compute the this leading divergence and then multiply by $i\pi \delta$. This is a great simplification because it dispenses us from computing the finite term $I_1(k)$ of the UV-divergent integral, which is in general much more complicated. Moreover, these manipulations already tell us that the parity-odd trispectrum is actually UV-finite! This is important because for the class of theories we consider here the tree-level contribution vanishes in general \cite{Liu:2019fag,Cabass:2022rhr} and so it would have been impossible to re-absorb the UV divergence into a counterterm.


\subsection{Momentum integrals}
The mode function of massless scalars and conformally coupled scalars can be written as derivative operators acting on a plane wave as in \eqref{onplane}. Hence, in general, we can recast $B_{4A}$ into the following form:
\begin{equation}
   B_{4A}=2\text{Im}\int_{\bfp}\delta^{(3)}(\bfp_1+\bfp_2+\bfs)\Tilde{F}(\{\bfk\},\bfp_1,\bfp_2)I_{\text{flat}},
\end{equation}
where $\Tilde{F}$ is a differential operator which depends on the form of the interactions $F_{\text{PE}}$ and $F_{\text{PO}}$, and 
\begin{multline}
    I_{\text{flat}}=\int_{-\infty}^{\eta_0} d\eta_1\int_{-\infty}^{\eta_0} d\eta_2 e^{i\omega_L\eta_1}e^{i\omega_R\eta_2}\left(e^{ip_1(\eta_2-\eta_1)}\theta(\eta_1-\eta_2)+e^{ip_1(\eta_1-\eta_2)}\theta(\eta_2-\eta_1)\right)\\\times\left(e^{ip_2(\eta_2-\eta_1)}\theta(\eta_1-\eta_2)+e^{ip_2(\eta_1-\eta_2)}\theta(\eta_2-\eta_1)\right),
\end{multline}
is a simpler integral involving only plane waves (hence the label "flat"). To simplify our notation we will define
\begin{align}
 \omega_L &=k_1+k_2\,, & \omega_R &=k_3+k_4 & k_T&=\sum_a^4 k_a=\omega_L+\omega_R. \,.
\end{align}
The time integral $I_\text{flat}$ gives:
\begin{equation}\label{Iflat}
    I_{\text{flat}}=\frac{1}{k_T}\left(\frac{1}{p_1+p_2+\omega_L}+\frac{1}{p_1+p_2+\omega_R}\right)\,.
\end{equation}
It will be convenient to change the integration measure of the momentum integral in the following way:
\begin{equation}
	\int\frac{d^{3+\delta}p}{(2\pi)^3}\delta^{(3)}(\bfp_1+\bfp_2+\bfs)f(\bfp)=\frac{1}{8\pi^2}\int_{s}^{\infty}dp_+\int_{-s}^{s}dp_-\frac{p_1p_2}{s}f(\bfp).
\end{equation}
where $p_+=p_1+p_2$ and $p_-=p_1-p_2$. After performing the $p_-$ integral, the remaining integral takes two possible forms. The first is 
\begin{equation}
    A_n=\int_{s}^{\infty}dp_+ (p_+)^{\delta+n} I_{\text{flat}}.
\end{equation}
As discussed above, we are only interested in the UV-divergent part of the integral. To find it, first note that:
\begin{equation}
    A_n=\int_{0}^{\infty}dp_+ (p_+)^{\delta+n} I_{\text{flat}}-\int_{0}^{s}dp_+ (p_+)^{\delta+n} I_{\text{flat}}
\end{equation}
For $\omega_L>0$ and $\omega_R>0$, the second integral is finite. The first integral can be written in terms of gamma functions, and can be simplified to give:
\begin{equation}
    A_n=\frac{(-1)^n}{\delta}\frac{\omega_L^n+\omega_R^n}{k_T}+(\text{finite}).
    \label{Anintegral}
\end{equation}
The second possible form of the $p_+$ integral is
\begin{equation}
    Z_n=\int^{\infty}_s dp_+(p_+)^{\delta+n}\log\left(\frac{p_++s}{p_+-s}\right)I_{\text{flat}}.
\end{equation}
At first glance, this integral looks like it has both UV and IR divergence. However when we evaluate the integral there is no IR divergence. This is because after we evaluate the integral, we get either $\log({p_+-s})$ multiplied by some power of $(p_+-s)$ (which is convergent) or dilogarithms which are not divergent. As an example, consider the case where $n=2$.  We obtain:
\begin{multline}
    Z_2=\frac{1}{2}\left(-(p_+-s)(p_++s-2\omega_L)\log(p_+-s)+(p_++s)(p_+-s-2\omega_L)\log(p_++s)\right)|_{p_+=s}\\
    +s(s+\omega_L)+\omega_L^2\left(\text{Li}_2\left(\frac{s-\omega_L}{s+\omega_L}\right)-\frac{\pi^2}{6}\right)+(\omega_L\rightarrow\omega_R)+\text{(UV divergent terms)}.
\end{multline}
This is not divergent after substituting $p_+=s$. A similar story applies to any integer $n$. 

To find the UV-divergent part of $Z_n$, we consider this integral instead:
\begin{equation}
    \frac{dZ_n}{ds}=\int_{s}^{\infty}dp_+p_+^{n+\delta}\left(\frac{1}{p_++s}+\frac{1}{p_+-s}\right)I_{\text{flat}}.
\end{equation}
This integral can be simplified by using partial fraction. Since this integral is not IR-divergent, we evaluate it in the same way as we did for $A_n$:
\begin{equation}
    \frac{dZ_n}{ds}=\frac{1}{k_T\delta}\left[\frac{(-\omega_L)^n-(-s)^n}{-\omega_L+s}+\frac{(-\omega_L)^n-s^n}{-\omega_L-s}+\frac{(-\omega_R)^n-(-s)^n}{-\omega_R+s}+\frac{(-\omega_R)^n-s^n}{-\omega_R-s}\right].
\end{equation}
This can be simplified into:
\begin{equation}
    \frac{dZ_n}{ds}=\frac{(-1)^{n-1}}{k_T\delta}\sum_{r=0}^{n-1}(\omega_L^{n-r-1}+\omega_R^{n-r-1})(s^r+(-s)^r).
\end{equation}
Therefore we obtain:
\begin{equation}\label{Zintegral}
    Z_n=\frac{(-1)^{n-1}}{k_T\delta}\sum_{r=0}^{n-1}\frac{1}{r+1}(\omega_L^{n-r-1}+\omega_R^{n-r-1})(s^{r+1}-(-s)^{r+1}).
\end{equation}

We can compute the out-of-time-ordered part of the trispectrum in a similar way: 
\begin{equation}
   B_{4B}=2\text{Im}\int_{\bfp}\delta^{(3)}(\bfp_1+\bfp_2+\bfs)\Tilde{F}(\{\bfk\},\bfp_1,\bfp_2)J_{\text{flat}},
\end{equation}
where
\begin{equation}
    J_{\text{flat}}=\int_{-\infty}^{\eta_0} d\eta_1\int_{-\infty}^{\eta_0} d\eta_2 e^{i\omega_L\eta_1}e^{-i\omega_R\eta_2}e^{i(p_1+p_2)(\eta_1-\eta_2)}.
\end{equation}
Evaluating the time integral gives us:
\begin{equation}
    J_{\text{flat}}=\frac{-1}{(p_1+p_2+\omega_L)(p_1+p_2+\omega_R)}.
\end{equation}
For the integral over the loop momentum we will encounter integrals again two different forms. The first is
\begin{equation}
    \tilde{A}_n=\int_{s}^{\infty}dp_+ (p_+)^{\delta+n} J_{\text{flat}}.
\end{equation}
Applying the same argument as we did for $A_n$, we obtain:
\begin{equation}
    \tilde{A}_n=\frac{(-1)^n}{\delta}\frac{\omega^n_L-\omega^n_R}{\omega_L-\omega_R}+(\text{finite})=\frac{(-1)^n}{\delta}\sum_{m=0}^{n-1}\omega_L^m\omega_R^{n-m-1}+(\text{finite}).\label{tilAintegral}
\end{equation}
The second possible form is
\begin{equation}
    \tilde{Z}_n=\int_{s}^{\infty}dp_+p^{n+\delta}_+\log\left(\frac{p_++s}{p_+-s}\right)J_{\text{flat}}.
\end{equation}
Applying the same argument as we did for $Z_n$, we obtain:
\begin{align}
    \label{tilZintegral}
    \tilde{Z}_n&=\frac{(-1)^{n-1}}{\delta(\omega_L-\omega_R)}\sum_{r=0}^{n-1}\frac{1}{r+1}(\omega_L^{n-r-1}-\omega_R^{n-r-1})(s^{r+1}-(-s)^{r+1})\\
    &=\nonumber \frac{(-1)^{n-1}}{\delta}\sum_{r=0}^{n-1}\frac{1}{r+1}\left(\sum_{m=0}^{n-r-2}\omega_L^{m}\omega_R^{n-r-2-m}\right)(s^{r+1}-(-s)^{r+1}).
\end{align}

By looking at \eqref{tilAintegral} and \eqref{tilZintegral}, we notice that if we take enough derivatives with respect to $\omega_L$ and $\omega_R$ we will get zero. For example, if we consider:
\begin{equation}\label{noB4B}
(\partial_{\omega_L})^{n_L}(\partial_{\omega_R})^{n_R}\tilde{A}_n,
\end{equation}
the result is zero for $n_L+n_R\geq n$. Similarly for $\tilde{Z}_n$, we will get zero if $n_L+n_R\geq n-1$. Since we need to take derivatives of these master integrals when computing the trispectrum, we will find that $B_{4B}$ vanishes.

Our general strategy for computing the trispectrum will be the following: write down the trispectrum as a differential operator acting on an integral, recasting the integral in terms of the master integrals $A_n$ and $Z_n$, then use our result from \eqref{Anintegral} and \eqref{Zintegral} to compute the divergent part and hence the final trispectrum.


\subsection{A toy model: $\sigma_1 \sigma_2 \rightarrow \sigma_a \sigma_b \rightarrow \sigma_3 \sigma_4$}

As a warm up example, let us consider a parity-odd trispectrum from the following interactions:
\begin{align}
\mathcal{L}_{PO}&=\epsilon^{ijk}\partial_i\sigma_1\partial_j\sigma_2\partial_k\sigma_a\sigma_b, &	\mathcal{L}_{PE}&=(\partial_i\sigma_3)\sigma_4(\partial^{i}\sigma_a)\sigma_b,
\end{align}
where all of the fields are conformally coupled scalars. Let's begin to consider the time-ordered part of the trispectrum $B_{4A}$. We have the integral:
\begin{multline}
	B_{4A}=2i\text{ Im}\int \frac{d^{3+\delta}p}{(2\pi)^3}\int \frac{d\eta_1}{(-H\eta_1)^{4+\delta}}\int \frac{d\eta_2}{(-H\eta_2)^{4+\delta}}(i^3(-H\eta_1)^3)(\bfk_1\times\bfk_2\cdot\bfp_1)\\\times(i^2(-H\eta_2)^2(\bfk_3\cdot\bfp_1))G_+(k_1,\eta_1)G_+(k_2,\eta_1)G_{++}(p_1,\eta_1,\eta_2)\\\times G_{++}(p_2,\eta_1,\eta_2)G_+(k_3,\eta_2)G_+(k_4,\eta_2).\label{toyB4A}
\end{multline}
First, we recast this integral in terms of an operator acting on a simpler integral. By counting powers of $\eta_1$ and $\eta_2$, we get:
\begin{equation}
	B_{4A}=2i\text{ Im}\frac{ (-H\eta_0)^4H^5}{64 k_1 k_2 k_3 k_4}(i\partial_{\omega_L})^{3+\delta}(i\partial_{\omega_R})^{2+\delta}\int\frac{d^{3+\delta}p}{(2\pi)^3}\frac{i^3(\bfk_1\times\bfk_2\cdot\bfp_1)i^2(\bfk_3\cdot\bfp_1)}{p_1p_2}I_{\text{flat}}.
\end{equation}
As argued in the previous section, we only need to compute:
\begin{equation}
B_{4A}=i\frac{ 2\pi\delta (H\eta_0)^4H^5}{64 k_1 k_2 k_3 k_4}(\bfk_1\times\bfk_2)^{i}(\bfk_3)^{j}(\partial_{\omega_L})^3(\partial_{\omega_R})^2\int\frac{d^{3+\delta}p}{(2\pi)^3}(\bfp_1)_i(\bfp_1)_j\frac{1}{p_1p_2}I_{\text{flat}}.
\end{equation}
Here we encounter the following tensorial integral
\begin{equation}
    I^{(2)}_{ij}=\int\frac{d^{3+\delta}p}{(2\pi)^3}(\bfp_1)_i(\bfp_1)_j\frac{1}{p_1p_2}I_{\text{flat}},
\end{equation}
which we can re-write as
\begin{equation}
    I^{(2)}_{ij}=I^{(2)}_0\delta_{ij}+\frac{s_is_j}{s^2}I^{(2)}_2.
\end{equation}
Since this integral is contracted with $\bfk_1\times\bfk_2$, the term $I^{(2)}_2$ does not contribute. Therefore we only need $I^{(2)}_0$, which is given by:
\begin{equation}
    I^{(2)}_0=\frac{1}{2}(\delta_{ij} I^{(2)}_{ij}+\frac{s_is_j}{s^2}I^{(2)}_{ij}).
\end{equation}
More explicitly, the integral is:
\begin{equation}
    I^{(2)}_0=\int \frac{d^{3+\delta}p}{(2\pi)^3}\frac{p_1^2-\frac{(\bfp_1\cdot\bfs)^2}{s^2}}{2p_1p_2}I_{\text{flat}}.
\end{equation}
This integral can be recast into the form:
\begin{equation}
    I^{(2)}_0=\frac{1}{8\pi^2}\frac{A_2-s^2A_0}{6}.
\end{equation}
Putting this back into $B_{4A}$, we have:
\begin{align}
\nonumber B_{4A}&=i\frac{ 2\pi (H\eta_0)^4H^5}{64 k_1 k_2 k_3 k_4}(\bfk_1\times\bfk_2)\cdot(\bfk_3)(\partial_{\omega_L})^3(\partial_{\omega_R})^2\left(\frac{1}{8\pi^2}\frac{A_2-s^2A_0}{6}\right)\\
&=-i\frac{(\bfk_1\times\bfk_2\cdot\bfk_3)H^9\eta_0^4}{64\pi k_1k_2k_3k_4 k_T^6}(\omega_L^2-6\omega_L\omega_R+3\omega_R^2-10s^2).\label{B4Atoy}
\end{align}

The procedure for computing $B_{4B}$ is similar, except we replace $I_{\text{flat}}$ with $J_{\text{flat}}$. This gives us:
\begin{equation}
   B_{4B}=i\frac{ 2\pi (H\eta_0)^4H^5}{64 k_1 k_2 k_3 k_4}(\bfk_1\times\bfk_2)\cdot(\bfk_3)(\partial_{\omega_L})^3(\partial_{\omega_R})^2\left(\frac{1}{8\pi^2}\frac{\tilde{A}_2-s^2\tilde{A}_0}{6}\right).
\end{equation}
Here we are taking $n_L=3$ derivatives with respect to $\omega_L$ and $n_R=2$ derivatives with respect to $\omega_R$. Since $n_L+n_R=5$ while the index of $\tilde A$ is 2 and 0, we expect $B_{4B}$ to vanish. This is confirmed by \eqref{tilAintegral}, which tells us that $\tilde{A}_0=0$ and $\tilde{A}_2=\frac{k_T}{\delta}$. Since $B_{4B}=0$, the only contribution to the trispectrum is $B_{4A}$.

In summary, the one-loop two-vertex parity-odd trispectrum is
\begin{align}\label{toy}
    \BPO=B_{4A}=-i\frac{(\bfk_1\times\bfk_2\cdot\bfk_3)H^9\eta_0^4}{64\pi k_1k_2k_3k_4 k_T^6}(\omega_L^2-6\omega_L\omega_R+3\omega_R^2-10s^2)\,.
\end{align}
A few comments are in order:
\begin{itemize}
    \item The result is UV finite, as anticipated around \eqref{ipidelta}. This is important because there is no tree-level counterterm to absorb this divergence. 
    \item The result has the expected scaling $B_4 \sim \eta_0^4/k^5$ for the trispectrum of a conformally coupled scalar, and is indeed parity odd because of the combination $\bfk_1\times\bfk_2\cdot\bfk_3$.
    \item The result is surprising simple: it is just a rational function in the momenta with the usual normalization $1/(k_1 k_2 k_3 k_4)$ and only a total-energy pole at $k_T=0$. This is the same structure as a tree-level contact diagram. The crucial difference is that $B_4$ cannot come from a contact wavefunction coefficient $\psi_4$ that obeys the cosmological optical theorem \cite{COT}. To see this, notice that at tree-level contact order, $B_4$ would need to come from a purely imaginary $\psi_4 \sim i k^3$, but then $\psi_4(k,\bfk)+\psi_4^\ast(-k,-\bfk)=2\psi_4(k,\bfk)\neq 0$. This check could be used to detect whether a given rational function arises or not as a contact diagram in a \textit{unitary} EFT. 
    \item Intriguingly $\BPO$ in \eqref{toy} could be attributed to a contact diagram in a \textit{non-unitary} EFT. Indeed the expression in \eqref{toy} contains the kinematic structures of a \textit{local} EFT, which were identified recently in \cite{Bonifacio:2021azc}. Non-unitary EFTs are expected to arise generically in open quantum systems. We will pursue this elsewhere.
    \item One may worry that our result is simply an artifacts of using dim reg. In appendix \ref{app:c} we compute $B_{4A}$ using cutoff regularization and find the same result.
\end{itemize}
We now move on to cases in which the external legs are massless scalars, which are more directly relevant for inflationary phenomenology.


\section{Conformally coupled loop}\label{sec:conformal}

In this section, we compute the contribution from the one-loop diagram in Figure \ref{fig:2siteloop}, where the four external legs correspond to a single massless scalar denoted by $\phi$. First, we show the result in the case in which the fields in the loop are two conformally coupled scalars $\sigma_a$ and $\sigma_b$. This represents a phenomenologically viable model of inflation with spectator massive fields. Second, we perform the calculation in single-clock inflation where all lines represent the same massless scalar $\phi$. 




\subsection{Conformally coupled spectator scalars: : $\phi \phi \to \sigma_a \sigma_b \to \phi \phi $}

Since we would like to consider a phenomenologically viable model, where the scalar field $\phi$ can be identified with the Goldstone boson $\pi$ of time-translations in the effective field theory of inflation \cite{Creminelli:2006xe,Cheung:2007st}, we consider interactions where $\phi$ has at least one time derivative, which would arise from $\delta g^{00}$, or two spatial derivative, which would arise from perturbations to the extrinsic curvature $K_{ij}$. For concreteness, we will consider the trispectrum from the following two interactions:
\begin{align}
\mathcal{L}_{PO}&=\epsilon_{ijk}\,\partial_{il}\phi\,\dot{\phi}\partial_{jl}\sigma_a\partial_k\sigma_b,\\
\mathcal{L}_{PE}&=\partial_{ij}\phi\,\partial_i\sigma_a\partial_j\dot{\phi}\sigma_b.
\end{align}
where $\sigma_{a,b}$ are conformally coupled and $\phi$ is massless. As in the previous section, we want to write down the corresponding integral as some differential operators acting on a simpler integral. The differential operator corresponding to the left vertex, which we choose to be the one with an odd number of spatial derivatives, is given by\footnote{Here we separated $-i\partial_{\omega_L}$, which accounts for the factor of $\eta$ in $\dot \phi$ from all other factors of $a^{-1}=-H\eta$ from $\sqrt{-g}$, derivatives and the conformally coupled mode functions, which are accounted for by $(-i\partial_{\omega_L})^4$.}
\begin{align}
\hat{L}=i^5\left[\bfk_1\cdot(\bfp_1\times\bfp_2)(\bfk_1\cdot\bfp_1)k_2^2 O^{(L)}_{k_1}+\bfk_2\cdot(\bfp_1\times\bfp_2)(\bfk_2\cdot\bfp_1)k_1^2O^{(L)}_{k_2}\right](-i\partial_{\omega_L})(-iH\partial_{\omega_L})^{6+2-4}\,,
\end{align}
where we defined $O^{(i)}_{k}=1-k\partial_{\omega_i}$. Notice that for both the conformally couple fields and for $\dot \phi$ we don't need a dedicated differential operator because the mode functions are already proportional to a plane wave and the overall factor of $\eta$ is captured by the $\partial_{\omega_L}$ operator. This can be simplified into:
\begin{equation}
	\hat{L}=H^4(\bfk_1\times\bfk_2)^{i}\left(k_2^2O^{(L)}_{k_1}\bfk_1^{j}-k_1^2O^{(L)}_{k_2}\bfk_2^{j}\right)(\partial_{\omega_L})^5(\bfp_1)_i(\bfp_1)_j
\end{equation}
Similarly for the right vertex, we have:
\begin{equation}
	\hat{R}=(\bfk_3\cdot\bfk_4)\left(k_4^2O_{k_3}^{(R)}\bfk_3^{i}+k_3^2O_{k_4}^{(R)}\bfk_4^{i}\right)(-iH\partial_{\omega_R})^{5+2-4}(-i\partial_{\omega_R})(\bfp_1)_i
\end{equation}
The trispectrum is:
\begin{multline}
	B_{4A}=(2\pi 
 i\delta)\frac{H^{19}}{16 k_1^3k_2^3k_3^3k_4^3}(\bfk_1\times\bfk_2)^{i}\left(k_2^2O^{(L)}_{k_1}\bfk_1^{j}-k_1^2O^{(L)}_{k_2}\bfk_2^{j}\right)\\\times(\bfk_3\cdot\bfk_4)\left(k_4^2O_{k_3}^{(R)}\bfk_3^{l}+k_3^2O_{k_4}^{(R)}\bfk_4^{l}\right)(\partial_{\omega_L})^5(\partial_{\omega_R})^4I^{(3)}_{ijl},
\end{multline}
where $I^{(3)}_{ijl}$ is given by:
\begin{equation}
I^{(3)}_{ijl}=\int\frac{d^3 p}{(2\pi)^3}\frac{(\bfp_1)_i(\bfp_1)_j(\bfp_1)_l}{4p_1p_2}I_{\text{flat}}.
\end{equation}
where $I_{\text{flat}}$ was given in \eqref{Iflat}. Once again we can separate the tensorial integral into scalar integrals (this can be done more systematically as discussed in Appendix \ref{app:B}):
\begin{equation}
I_{ijl}^{(3)}=\left(\frac{I^{(3)}_1}{s}(s_i\delta_{jk}+s_j\delta_{ik}+s_k\delta_{ij})+\frac{I^{(3)}_3}{s^3}s_is_js_k\right).
\end{equation}
Notice that this expression assumes $s=|\mathbf{s}|\neq 0$, but it otherwise does not depend on $s$. For our purposes, we only need $I_1$ because the other terms vanish once contracted with the epsilon tensor. This is given by
\begin{equation}
	I^{(3)}_1=\frac{1}{2}\left(\frac{1}{s}s_i\delta_{jl}I^{(3)}_{ijl}-\frac{1}{s^3}s_is_js_lI^{(3)}_{ijl}\right).
\end{equation}
More explicitly, we have:
\begin{equation}
I^{(3)}_1=\int_{\bfp}\frac{s^2p_1^2(\bfp_1\cdot\bfs)-(\bfp_1\cdot\bfs)^3}{s^3p_1p_2}I_{\text{flat}}.
\end{equation}
Computing this integral gives us:
\begin{multline}
B_{4A}=(2\pi i\delta)\frac{H^{19}}{16 k_1^3k_2^3k_3^3k_4^3}\left(k_2^2O^{(L)}_{k_1}(\bfk_1\cdot\bfs)-k_1^2O^{(L)}_{k_2}(\bfk_2\cdot\bfs)\right)(\bfk_3\cdot\bfk_4)\\\times\left(k_4^2O_{k_3}^{(R)}\bfk_3\cdot(\bfk_1\times\bfk_2)+k_3^2O_{k_4}^{(R)}\bfk_4\cdot(\bfk_1\times\bfk_2)\right)(\partial_{\omega_L})^5(\partial_{\omega_R})^4 \frac{1}{32\pi^2}\frac{A_2-s^2A_0}{12}.
\end{multline}
Using our general results for the $A$ integrals, this can be further simplified into:
\begin{multline}\label{finalss}
B_{4A}=i\frac{H^{19}}{16 k_1^3k_2^3k_3^3k_4^3}\left[k_2^2O^{(L)}_{k_1}(\bfk_1\cdot\bfs)-k_1^2O^{(L)}_{k_2}(\bfk_2\cdot\bfs)\right](\bfk_3\cdot\bfk_4)\\\times\left[k_4^2O_{k_3}^{(R)}\bfk_3\cdot(\bfk_1\times\bfk_2)+k_3^2O_{k_4}^{(R)}\bfk_4\cdot(\bfk_1\times\bfk_2)\right]\left[\frac{1}{16\pi}\frac{3360(18s^2-3\omega_L^2+10\omega_L\omega_R-5\omega_R^2)}{k_T^{10}}\right].
\end{multline} 

Since $O^{(L)}_k$ and $O^{(R)}_k$ each provides an extra derivative, $B_{4A}$ has a $k_T$ pole of order $12$. This matches with the standard expectation that the order $p$ of the $k_T$ pole is \cite{BBBB}
\begin{equation}
    p=1+\sum_i(\Delta_i-4)=1+(10-4)+(9-4)=12.
\end{equation}

Similarly, we can compute $B_{4B}$:
\begin{multline}
B_{4B}=(2\pi i\delta)\frac{H^{19}}{16 k_1^3k_2^3k_3^3k_4^3}\left(k_2^2O^{(L)}_{k_1}(\bfk_1\cdot\bfs)-k_1^2O^{(L)}_{k_2}(\bfk_2\cdot\bfs)\right)(\bfk_3\cdot\bfk_4)\\\times\left(k_4^2O_{k_3}^{(R)}\bfk_3\cdot(\bfk_1\times\bfk_2)+k_3^2O_{k_4}^{(R)}\bfk_4\cdot(\bfk_1\times\bfk_2)\right)(\partial_{\omega_L})^5(\partial_{\omega_R})^4 \frac{1}{32\pi^2}\frac{\tilde{A}_2-s^2\tilde{A}_0}{12}.
\end{multline}
Since $n_L=5$, $n_R=4$ and $n\geq 2$, we have $n_L+n_R>n$, so when we take derivatives we find $B_{4B}=0$, as anticipated. In summary the final result is $\BPO=B_{4A}$ as given in \eqref{finalss}. The same remark as at the end of the previous section apply to this surprisingly simple result as well.


\subsection{Same internal fields}

Next we want to consider the case in which there is a single spectator scalar, with a conformally coupled mass. If we simply replace $\sigma_1=\sigma_2=\sigma$ in the above example, the spatial derivatives for the internal fields in the parity even vertex can all be removed by integration by parts. As a result the trispectrum vanishes. Instead, by direct investigation we found that the following interactions provide a non-vanishing result that is minimal in terms of number of derivatives:
\begin{align}
\mathcal{L}_{PO}&=\epsilon_{ijk}\,\partial_{il}\phi\,\dot{\phi}\partial_{jl}\sigma\partial_k\sigma,\\
\mathcal{L}_{PE}&=\partial_{ij}\phi\,\partial_{ij}\sigma\dot{\phi}\sigma.
\end{align}

Consider the left vertex first. We need the differential operator
\begin{multline}
	\hat{L}=iH^4(\bfk_1\times\bfk_2)\cdot\bfp_1\left(k_2^2O^{(L)}_{k_1}\bfk_1\cdot\bfp_1-k_1^2O^{(L)}_{k_2}\bfk_2\cdot\bfp_1\right)(\partial_{\omega_L})^5\\
	+iH^4(\bfk_1\times\bfk_2)\cdot\bfp_2\left(k_2^2O^{(L)}_{k_1}\bfk_1\cdot\bfp_2-k_1^2O^{(L)}_{k_2}\bfk_2\cdot\bfp_2\right)(\partial_{\omega_L})^5
\end{multline}
Since $(\bfk_1\times\bfk_2)\cdot\bfp_1=(\bfk_1\times\bfk_2)\cdot(-\bfs-\bfp_1)=-(\bfk_1\times\bfk_2)\cdot\bfp_2$, this simplifies to
\begin{equation}
	\hat{L}=iH^4(\bfk_1\times\bfk_2)^{i}\left(k_2^2O^{(L)}_{k_1}\bfk_1^{j}-k_1^2O^{(L)}_{k_2}\bfk_2^{j}\right)(\partial_{\omega_L})^5(\bfp_1)_i[(\bfp_1)_j-(\bfp_2)_j]\,.
\end{equation}
The right vertex gives:
\begin{equation}
	\hat{R}=H^3\left((\bfk_3\cdot\bfp_1)^2O^{(R)}_{k_3}k_4^2+(\bfk_4\cdot\bfp_1)^2O^{(R)}_{k_4}k_3^2\right)(\partial_{\omega_R})^4+(\bfp_1\rightarrow\bfp_2).
\end{equation}
So the trispectrum is found to be
\begin{multline}
	B_{4A}=(2\pi 
 i\delta)\frac{H^{19}}{16 k_1^3k_2^3k_3^3k_4^3}(\bfk_1\times\bfk_2)^{i}\left(k_2^2O^{(L)}_{k_1}\bfk_1^{j}-k_1^2O^{(L)}_{k_2}\bfk_2^{j}\right)\\ \times\left(k_4^2O_{k_3}^{(2)}\bfk_3^{l}\bfk_3^{m}+k_3^2O_{k_4}^{(2)}\bfk_4^{l}\bfk_4^{m}\right)(\partial_{\omega_L})^5(\partial_{\omega_R})^4I_{ijlm},
\end{multline}
where
\begin{multline}
	I_{ijlm}=\int_{\bfp}\frac{1}{4p_1p_2}\left((\bfp_1)_i(\bfp_1)_j(\bfp_1)_l(\bfp_1)_m+(\bfp_1)_i(\bfp_1)_j(\bfp_2)_l(\bfp_2)_m\right.\\-\left.(\bfp_1)_i(\bfp_2)_j(\bfp_1)_l(\bfp_1)_m-(\bfp_1)_i(\bfp_2)_j(\bfp_2)_l(\bfp_2)_m \right) I_{\text{flat}}.
\end{multline}
Since we can always exchange $(\bfp_1)_i$ for $-(\bfp_2)_i$ (as it is contracted with $\bfk_1\times\bfk_2$), and we can also exchange $\bfp_1$ for $\bfp_2$ by changing the integration variable from $\bfp_1$ to $-\bfp_1-\bfs$, the integral simplifies into:
\begin{equation}
    I_{ijlm}=\int_{\bfp}\frac{4(\bfp_1)_i(\bfp_1)_j(\bfp_1)_l(\bfp_1)_m+2(\bfp_1)_i(\bfs)_j(\bfp_1)_l(\bfp_1)_m}{4p_1p_2}\,.
\end{equation}

With this, we can use the tensor structure results in Appendix \ref{app:B} to compute the tensorial integral in terms of scalar integrals. This yields:
\begin{align}
    I_{ijlm}=&\frac{1}{16\pi^2}\frac{1}{30}\left[\left(\delta_{il}\delta_{jm}+\delta_{im}\delta_{jl}\right)(A_4-2s^2A_2+s^4A_0)\right.\nonumber\\&+\left.\left(\delta_{il}s_js_m+\delta_{im}s_js_l\right)(A_2-s^2A_0)\right].\label{Iijlm1}
\end{align}
Putting this back in the trispectrum, we obtain:
\begin{align}\label{finalss2}
    B_{4A}=&i\frac{H^{19}}{16 k_1^3k_2^3k_3^3k_4^3}\left[\left(k_2^2O^{(L)}_{k_1}\bfk_1-k_1^2O^{(L)}_{k_2}\bfk_2\right)\cdot\left(k_4^2O_{k_3}^{(2)}(\bfk_1\times\bfk_2\cdot\bfk_3)\bfk_3+k_3^2O_{k_4}^{(2)} \nonumber(\bfk_1\times\bfk_2\cdot\bfk_4)\bfk_4\right)\right.\\
    &\times \nonumber \frac{-1}{8\pi}\frac{192(126s^4+\omega_L^4-20\omega_L^3\omega_R+60\omega_L^2\omega_R^2-40\omega_L\omega_R^3+5\omega_R^4-14s^2(3\omega_L^2-10\omega_L\omega_R+5\omega_R^2))}{k_T^{10}}\\
    &+\left(k_2^2O^{(L)}_{k_1}(\bfk_1\cdot\bfs)-k_1^2O^{(L)}_{k_2}(\bfk_2\cdot\bfs)\right)\left(k_4^2O_{k_3}^{(2)}(\bfk_1\times\bfk_2\cdot\bfk_3) \nonumber(\bfk_3\cdot\bfs)+k_3^2O_{k_4}^{(2)}(\bfk_1\times\bfk_2\cdot\bfk_4)(\bfk_4\cdot\bfs)\right)\\
    &\times\left. \frac{1}{8\pi}\frac{1344(18s^2-3\omega_L^2+10\omega_L\omega_R-4\omega_R^2)}{k_T^{10}}\right].
\end{align}

To compute $B_{4B}$, we just need to replace $A_n$ in \eqref{Iijlm1} with $\tilde{A}_n$. But once again $n_L=5$, $n_R=4$ and $n\geq 4$, so $n_L+n_R>n$ and we have $B_{4B}=0$. 

\paragraph{Permutations}

One also need to sum over permutations when calculating the correlator. For instance, it is necessary to also consider the correlator when the left and right verticies are swapped. Notice that the polynomials and the operators in the final result above are not symmetric under the exchange of $\omega_L$ and $\omega_R$, hence the correlator does not vanish upon summing over permutations. 

Similarly one may worry whether summing over the $(s,t,u)$ channels may result in some cancellation. However, when considering these permutations, one also needs to redefine $\omega_L$ and $\omega_R$: for the $t-$channel, $\omega_L=k_1+k_3$ while for the $u-$channel, $\omega_L=k_1+k_4$. In general, summing over different channels does not result in cancellation of the correlator as well.


\section{Massless loop: $\phi \phi \to \phi \phi \to \phi \phi $}\label{sec:single}

We now consider the case of single-clock inflation, where all lines represent a massless scalar $\phi$, to be identified with the Goldstone boson $\pi$ of time translations in the EFT of inflation. Conceptually the calculation is just the same as in previous examples. However the main new difficulty is to find interactions that give a non-vanishing result when symmetrised. The idea is that we have to include a sufficient number of derivatives such that all $\phi$'s appearing in the parity-odd interaction are distinct from each other. Furthermore, we also need a sufficient number of derivatives in the parity-even interactions to ensure that, after the loop integral has been computed, a term of the form $\bfk_1\times\bfk_2 \cdot \bfk_3$ can be generated. This results in a large number of derivatives and hence an algebraically more complex result, but no new conceptual issue emerges.

A minimal choice of interactions that gives a non-vanishing result is
\begin{align}
   \label{leading operator PO}
	&\mathcal{L}_{PO}=\lambda_{PO}\epsilon_{ijk}\partial_m\partial_n\phi\,\partial_n\partial_i\phi\,\partial_m\partial_l\partial_j\phi\,\partial_l\partial_k\phi,\\
    \label{leading operator PE}
	&\mathcal{L}_{PE}=\lambda_{PE}\dot{\phi}^2(\partial_i\partial_j\phi)^2
\end{align}
Let's follow a by now familiar script and start building the relevant differential operators. For the parity-odd interaction, the tensor structure of the vertex looks like:
\begin{equation}
	F_{PO}(p_1,p_2,k_1,k_2)=\frac{i^9}{a^{9}} \bfk_2\cdot(\bfp_1\times\bfp_2)(\bfk_1\cdot\bfp_1)(\bfk_1\cdot\bfk_2)(\bfp_1\cdot\bfp_2)+\text{permutations} 
\end{equation}
Notice that since $\bfp_1+\bfp_2+\bfk_1+\bfk_2=0$, we can always rearrange the cross product to take the form $\pm\bfk_2\cdot(\bfp_1\times\bfp_2)$. Carefully considering all the permutations, we obtain:
\begin{align}
	\nonumber(-ia)^9F_{PO}&=-2(\bfk_1\times\bfk_2)\cdot\bfp_1\left\{ (\bfk_1\cdot\bfk_2)(\bfp_1\cdot\bfp_2)(\bfk_1-\bfk_2)\cdot(\bfp_1-\bfp_2)\right.\\
	\nonumber&+(\bfp_2\cdot\bfk_1)(\bfp_2\cdot\bfk_2)\bfp_1\cdot(\bfk_1-\bfk_2)-(\bfp_1\cdot\bfk_1)(\bfp_1\cdot\bfk_2)\bfp_2\cdot(\bfk_1-\bfk_2)\\ &\left.+\left[(\bfk_1\cdot\bfp_2)(\bfk_2\cdot\bfp_1)-(\bfp_1\cdot\bfk_1)(\bfk_2\cdot\bfp_2)(\bfp_1\cdot\bfp_2+\bfk_1\cdot\bfk_2)\right]\right\}
\end{align} 
Note that $\bfp_1$ and $\bfp_2$ can be exchanged by changing the integration variable. Since we also sum over permutations on the parity-even vertex as well, we have:
\begin{align}
	\nonumber(-ia)^9V_{PO}&=-2(\bfk_1\times\bfk_2)\cdot\bfp_1\left[2(\bfk_1\cdot\bfk_2)(\bfp_1\cdot\bfp_2)(\bfp_1\cdot\bfr)-2(\bfk_1\cdot\bfp_1)(\bfk_2\cdot\bfp_1)(\bfp_2\cdot\bfr)\right.\\&\left.-((\bfk_1\cdot\bfs)(\bfp_1\cdot\bfk_2)-(\bfk_2\cdot\bfs)(\bfp_1\cdot\bfk_1))(\bfp_1\cdot\bfp_2+\bfk_1\cdot\bfk_2)\right].
\end{align}

Here I defined $\bfr=\bfk_1-\bfk_2$. Now we can use $\bfp_1\cdot\bfp_2=\frac{1}{2}(s^2-p_1^2-p_2^2)$ to simplify this further.

The tensor structure for the parity-even vertex is straightforward to obtain. Note that only one internal line can have a spatial derivative, otherwise it can be shown that the integral gives us zero. The trispectrum can now be written as:
\begin{equation}
	B_{4A}=(2\pi i\delta)\frac{\lambda_{PO}\lambda_{PE}H^{8}}{8k_1^3k_2^3k_3^3k_4^3}\int_\bfp \hat{L}\hat{R}\frac{H^4}{4p_1^3p_2^3}I_{\text{flat}}.
\end{equation}
The left operator is given by:
\begin{align}
	\hat{L}&=(L^{(2a)}+L^{(2b)}+L^{(4)})(-iH\partial_{\omega_L})^5O^{(L)}_{k_1}O^{(L)}_{k_2}O^{(L)}_{p_1}O^{(L)}_{p_2},\\
	L^{(2a)}&=-2(\bfk_1\times\bfk_2)^{i}\left[\frac{3}{2}(\bfk_1\cdot\bfk_2)\bfr^{j}-\frac{1}{2}(k_1^2\bfk_2^{j}-k_2^2\bfk_1^{j})\right](s^2-p_1^2-p_2^2)(\bfp_1)_i(\bfp_1)_j,\\
	L^{(2b)}&=-2(\bfk_1\times\bfk_2)^{i}\left[(\bfk_1\cdot\bfk_2)^2\bfr^{j}-(\bfk_1\cdot\bfk_2)(k_1^2\bfk_2^{j}-k_2^2\bfk_1^{j})\right](\bfp_1)_i(\bfp_1)_j,\\
	L^{(4)}&=4(\bfk_1\times\bfk_2)^{i}(\bfk_1)^{j}(\bfk_2)^{l}(\bfr)^{m}(\bfp_1)_i(\bfp_1)_j(\bfp_1)_l(\bfp_2)_m.
\end{align}
The right vertex becomes:
\begin{equation}
	\hat{R}=2\left((\bfk_3)^{i}(\bfk_3)^{j}k_4^2O^{(R)}_{k_3}O^{(R)}_{p_1}+(\bfk_4)^{i}(\bfk_4)^{j}k_3^2O^{(R)}_{k_4}O^{(R)}_{p_1}\right)p_2^2(\bfp_1)_i(\bfp_1)_j(-iH\partial_{\omega_R})^2(-i\partial_{\omega_R})^2
\end{equation}

Let us separate the trispectrum into three terms, where each term corresponds to one of the operators above:
\begin{equation}\label{finalBPO}
	B_{4A}=B_{4A}^{2a}+B_{4A}^{2b}+B_{4A}^{4}.
\end{equation}
The first term is
\begin{multline}
	B_{4A}^{2a}=(2\pi i)\frac{\lambda_{PO}\lambda_{PE}H^{14}}{8k_1^3k_2^3k_3^3k_4^3} (-2) (\bfk_1\times\bfk_2)^{i}\left[\frac{3}{2}(\bfk_1\cdot\bfk_2)\bfr^{j}-\frac{1}{2}(k_1^2\bfk_2^{j}-k_2^2\bfk_1^{j})\right] \\
 \times 2\left((\bfk_3)^{l}(\bfk_3)^{m}k_4^2O^{(R)}_{k_3}+(\bfk_4)^{l}(\bfk_4)^{m}k_3^2O^{(R)}_{k_4}\right) (-iH\partial_{\omega_L})^5(-i\partial_{\omega_R})^4O^{(L)}_{k_1}O^{(L)}_{k_2}I^{2a}_{ijlm},
\end{multline}
where 
\begin{equation}
	I_{ijlm}^{2a}:=\frac{1}{8}(\delta_{il}\delta_{jm}+\delta_{im}\delta_{jl})I^{2a}_{0}+\frac{1}{8s^2}(\delta_{il}s_js_m+\delta_{im}s_js_l)I^{2a}_{2}.
\end{equation}

Similarly, the second term is
\begin{multline}
	B_{4A}^{2b}=(2\pi i)\frac{\lambda_{PO}\lambda_{PE}H^{14}}{8k_1^3k_2^3k_3^3k_4^3} (-2)(\bfk_1\times\bfk_2)^{i}\left[(\bfk_1\cdot\bfk_2)\bfr^{j}-(k_1^2\bfk_2^{j}-k_2^2\bfk_1^{j})\right] (\bfk_1\cdot\bfk_2)\\
 \times  2\left((\bfk_3)^{l}(\bfk_3)^{m}k_4^2O^{(R)}_{k_3}+(\bfk_4)^{l}(\bfk_4)^{m}k_3^2O^{(R)}_{k_4}\right) (-iH\partial_{\omega_L})^5(-i\partial_{\omega_R})^4O^{(L)}_{k_1}O^{(L)}_{k_2}I^{2b}_{ijlm},
\end{multline}
where 
\begin{align}
    I_{ijlm}^{2b}
	&:=\frac{1}{8}(\delta_{il}\delta_{jm}+\delta_{im}\delta_{jl})I^{2b}_{0}+\frac{1}{8s^2}(\delta_{il}s_js_m+\delta_{im}s_js_l)I^{2b}_{2}
\end{align}
The $B^{4}_{4A}$ integral can be written similarly:
\begin{multline}
	B^{4}_{4A}=(2\pi i)\frac{\lambda_{PO}\lambda_{PE}H^{14}}{8k_1^3k_2^3k_3^3k_4^3} 4(\bfk_1\times\bfk_2)^{i}(\bfk_1)^{j}(\bfk_2)^{k}(\bfr)^{l} \\
 \times 2\left((\bfk_3)^{m}(\bfk_3)^{n}k_4^2O^{(R)}_{k_3}+(\bfk_4)^{m}(\bfk_4)^{n}k_3^2O^{(R)}_{k_4}\right) (-iH\partial_{\omega_L})^5(-i\partial_{\omega_R})^4O^{(L)}_{k_1}O^{(L)}_{k_2}I^{4}_{ijklmn},
\end{multline}
where
\begin{align}
	\nonumber I^{4}_{ijklmn}&=\frac{1}{57}\left(\delta_{im}(\delta_{jk}\delta_{ln}+\text{perm})+\delta_{in}(\delta_{jk}\delta_{lm}+\text{perm})\right)I^{4a}_0\\
	\nonumber&+\frac{1}{456}\left(\frac{\delta_{im}}{s^2}(s_js_k\delta_{ln}+\text{perm})+\frac{\delta_{in}}{s^2}(s_js_k\delta_{lm}+\text{perm})\right)I^{4a}_2\\
	\nonumber&+\frac{1}{76}\left(\frac{\delta_{im}}{s^4}s_js_ks_ls_n+\frac{\delta_{in}}{s^4}s_js_ks_ls_m\right)I^{4a}_4\\
	\nonumber&+\frac{1}{8}\left(\frac{\delta_{im}}{s^2}(\delta_{jk}s_ls_n+\text{perm})+\frac{\delta_{in}}{s^2}(\delta_{jk}s_ls_m+\text{perm})\right)I^{4b}_1\\
	&+\frac{1}{8}\left(\frac{\delta_{im}}{s^4}(s_js_ks_ls_n)+\frac{\delta_{in}}{s^2}(s_js_ks_ls_m)\right)I^{4b}_3.
\end{align}
Each of these terms can be obtained by using the tensor structure formula derived in Appendix \ref{app:B}. Then, we using our general strategy of rewriting integrals in terms of the $A_n$ and $Z_n$ master integrals. Eventually, it can be shown that:
\begin{align}
	\nonumber I^{2a}_0&=\frac{1}{27720 (\omega_L+\omega_R)^3}\left[3404 s^6-44 s^4 \left(142 \omega_L^2+106 \omega_L \omega_R+221 \omega_R^2\right)\right.\\\nonumber&+33 s^2 \left(201 \omega_L^4+623 \omega_L^3 \omega_R+855 \omega_L^2 \omega_R^2+229 \omega_L \omega_R^3+104 \omega_R^4\right)\\&-\left.231 \left(20 \omega_L^6+60 \omega_L^5 \omega_R+51 \omega_L^4 \omega_R^2+3 \omega_L^3 \omega_R^3+11 \omega_L^2 \omega_R^4+13 \omega_L \omega_R^5+6 \omega_R^6\right)\right]\\I^{2a}_2&=\frac{-7784 s^6+88 s^4 \left(83 \omega_L^2+38 \omega_L \omega_R+121 \omega_R^2\right)-33 s^2 \left(51 \omega_L^4+203 \omega_L^3 \omega_R+391 \omega_L^2 \omega_R^2+121 \omega_L \omega_R^3+50 \omega_R^4\right)}{13860 (\omega_L+\omega_R)^3}\\
	I^{2b}_0&=\frac{172 s^4-24 s^2 \left(11 \omega_L^2+9 \omega_L \omega_R+17 \omega_R^2\right)+21 \left(9 \omega_L^4+27 \omega_L^3 \omega_R+35 \omega_L^2 \omega_R^2+9 \omega_L \omega_R^3+4 \omega_R^4\right)}{1260 (\omega_L+\omega_R)^3}\\
	\nonumber I^{2b}_2&=\frac{2 s^2 \left(-97 s^2+60 \omega_L^2+30 \omega_L \omega_R+87 \omega_R^2\right)}{315 (\omega_L+\omega_R)^3}
\end{align}
\begin{align}
	\nonumber I^{4a}_0&=\frac{1}{360360 (\omega_L+\omega_R)^3}\left[-10212 s^6+156 s^4 \left(142 \omega_L^2+106 \omega_L \omega_R+221 \omega_R^2\right)\right.\\&\nonumber-143 s^2 \left(201 \omega_L^4+623 \omega_L^3 \omega_R+855 \omega_L^2 \omega_R^2+229 \omega_L \omega_R^3+104 \omega_R^4\right)\\&+\left.1287 \left(20 \omega_L^6+60 \omega_L^5 \omega_R+51 \omega_L^4 \omega_R^2+3 \omega_L^3 \omega_R^3+11 \omega_L^2 \omega_R^4+13 \omega_L \omega_R^5+6 \omega_R^6\right)\right]\\
	\nonumber I^{4a}_{2}&=\frac{1}{720720 (\omega_L+\omega_R)^3}\left[1168476 s^6-156 s^4 \left(8830 \omega_L^2+4366 \omega_L \omega_R+12983 \omega_R^2\right)\right.\\\nonumber&+143 s^2 \left(2865 \omega_L^4+11075 \omega_L^3 \omega_R+20691 \omega_L^2 \omega_R^2+6337 \omega_L \omega_R^3+2636 \omega_R^4\right)\\&\left.+11583 \left(20 \omega_L^6+60 \omega_L^5 \omega_R+51 \omega_L^4 \omega_R^2+3 \omega_L^3 \omega_R^3+11 \omega_L^2 \omega_R^4+13 \omega_L \omega_R^5+6 \omega_R^6\right)\right]\\
	\nonumber I^{4a}_{4}&=\frac{1}{480480 (\omega_L+\omega_R)^3}\left[-1122380 s^6+52 s^4 \left(10638 \omega_L^2+2666 \omega_L \omega_R+14013 \omega_R^2\right)\right.\\\nonumber&+715 s^2 \left(201 \omega_L^4+623 \omega_L^3 \omega_R+855 \omega_L^2 \omega_R^2+229 \omega_L \omega_R^3+104 \omega_R^4\right)\\&\left.-6435 \left(20 \omega_L^6+60 \omega_L^5 \omega_R+51 \omega_L^4 \omega_R^2+3 \omega_L^3 \omega_R^3+11 \omega_L^2 \omega_R^4+13 \omega_L \omega_R^5+6 \omega_R^6\right)\right]\\
	 I^{4b}_{1}&=\frac{-818 s^6+44 s^4 \left(25 \omega_L^2+16 \omega_L \omega_R+38 \omega_R^2\right)-33 s^2 \left(18 \omega_L^4+59 \omega_L^3 \omega_R+89 \omega_L^2 \omega_R^2+25 \omega_L \omega_R^3+11 \omega_R^4\right)}{13860 (\omega_L+\omega_R)^3}\\
	 I^{4b}_{3}&=-\frac{2 s^4 \left(-94 s^2+55 \omega_L^2+22 \omega_L \omega_R+77 \omega_R^2\right)}{495 (\omega_L+\omega_R)^3}\label{last}
\end{align}

To obtain $B_{4B}$, we simply replace $A_n$ with $\tilde{A}_n$ and $Z_n$ with $\tilde{Z}_n$. However, in the calculation, we find that $n\geq6$, while $n_L\geq5$ and $n_R\geq 4$. Hence when we take derivatives, we find $B_{4B}=0$ and so only $B_{4A}$ contributes to the two-vertex one-loop trispectrum.


\section{Signal-to-noise estimates}\label{sec:signal}

We have computed a one-loop contribution to the parity-odd trispectrum $ \BPO$. This can be important for observations because of the no-go theorems in \cite{Liu:2019fag,Cabass:2022rhr}. In particular, those results imply that for any number of scalar fields \textit{in the scale invariant limit} the parity-odd trispectrum of curvature perturbations vanishes at tree level. Therefore, the leading $ \BPO$ arises at one-loop order, as we computed in the previous sections. This is an exciting scenario because usually loop corrections are under control only when they are smaller than a corresponding tree-level contribution and hence cannot be expected to be the leading signal (exceptions include the single field chaotic inflationary models, see \cite{Abolhasani:2019lwu, Abolhasani:2020xcg}). For $\BPO$ things are different: it constitutes a rare example of a cosmological observable that probes loop contributions and can be in principle as large as allowed by the data. In the following, we will assume we are in the scale invariant limit of single-clock inflation and study how large the one-loop $ \BPO$ can be. For a cosmic variance limited experiment, we find that the signal-to-noise ratio in the tree-level parity-even trispectrum is always larger than that in the parity-odd trispectrum. However, we also point out that parity-even and parity-odd sectors can have different noise levels and different systematics. Hence, for certain datasets, it can make sense to search for our predicted signal $ \BPO$ in the data. In this section, we do not invoke technical naturalness. Our conclusions apply to any EFT, whether it is natural or not. In the next section we will discuss the additional constraints from naturalness. \\

The signal-to-noise ratio $S/N $ is an estimate of when a signal becomes observable, which happens when $ S/N >1$. To be as general as possible we don't commit to a specific observable. Instead we assume we can measure the profile of $ \phi(\bfk)$ in some volume $ V \sim \kmin^{-3} $ with a resolution $ \kmax >\kmin$. For an $ n$-point function the signal-to-noise ratio is
\begin{align}\label{SN}
\left(  \frac{S}{N}\right)^{2}=V^{n}\int_{\bfk_{1}\dots\bfk_{n}}\frac{\ex{\prod_{a}^{n}\phi(\bfk_{a})}\ex{\prod_{a}^{n}\phi(\bfk_{a})}}{\ex{\prod_{a}^{n}\phi(\bfk_{a})^{2}}}\,.
\end{align}
Notice that this is independent of the normalization of the field $\phi$. Therefore, without loss of generality, we will proceed assuming that $\phi$ is a canonically normalized scalar. To begin with, we assume that the speed of sound equals the speed of light, $c_s=c=1$, but we consider a more general speed of sound at the end this section. Let's now consider a model with the following two interactions,
\begin{align}\label{lint}
H_{\text{int}}=\frac{1}{\Lambda_{\text{PO}}^{9}}\partial_{i}^{9}\phi^{4}+\frac{1}{\Lambda_{\text{PE}}^{6}}\partial_{\mu}^{6}\phi^{4}\,,
\end{align}
where $ \partial_{i}^{9}$ denotes some unspecified contraction of nine spatial derivatives and $ \partial_{\mu}^{6}$ that of six temporal or spatial derivatives. Here $ \Lambda_{\text{PE,PO}}$ are the scales suppressing the respective higher-dimensional interactions. Then the parity-odd trispectrum appears first at one-loop order using both the parity-odd and parity-even interactions. A rough estimate is 
\begin{align}\label{bpo}
\BPO\sim\left(  \frac{H}{\Lambda_{\text{PO}}}\right)^{9}\left(  \frac{H}{\Lambda_{\text{PE}}}\right)^{6}\frac{H^{4}}{k^{9}} \frac{1}{16\pi^{2}}\,.
\end{align}
where we used scale invariance to conclude that $ B_{4}\sim k^{-9}$ and the factor of $ 1/(4\pi)^{2}$ is expected for a one-loop diagram. This term crucially relies on the existence of a non-vanishing parity-even quartic vertex and hence is necessarily associated with a tree-level contact parity-even trispectrum, which is generated by a single instance of the second interaction in \eqref{lint}. A  rough estimate of the resulting tree-level parity-even trispectrum is
\begin{align}\label{bpe}
\BPE\sim \left(  \frac{H}{\Lambda_{PE}}\right)^{6}\frac{H^{4}}{k^{9}}\,.
\end{align}
Using \eqref{SN}, the $ S/N$ for the one-loop parity-odd trispectrum is estimated to be (using $ (2\pi)^{3}\delta^{(3)}(\mathbf{0})=V$ and dropping numerical factors)
\begin{align}
\left( \frac{S}{N} \right)_{PO}\simeq \left(  \frac{H}{\Lambda_{\text{PO}}}\right)^{9}\left(  \frac{H}{\Lambda_{\text{PE}}}\right)^{6}\left( \frac{\kmin}{\kmax} \right)^{3/2} \frac{1}{(2\pi)^{9}} \frac{1}{16\pi^{2}}\,.
\end{align}
Similarly, the $ S/N$ for the parity-even trispectrum is found to be
\begin{align}
\left( \frac{S}{N} \right)_{\text{PE}}\simeq \left(  \frac{H}{\Lambda_{\text{PE}}}\right)^{6}\left( \frac{\kmin}{\kmax} \right)^{3/2} \frac{1}{(2\pi)^{9}}\,.
\end{align}
Taking their ratio we discover that
\begin{align}
\frac{\left(  S/N\right)_{\text{PO}}}{\left(  S/N\right)_{\text{PE}}}=\left(  \frac{H}{\Lambda_{\text{PO}}}\right)^{9}\frac{1}{16\pi^{2}}\ll 1\,.
\end{align}
This inequality tells us that the best chance to first see a signal in the parity-odd trispectrum arises when we take $ \Lambda_{\text{PO}}$ dangerously close to $ H$. Provided that the instrumental noise in the parity-even and odd measurement is comparable (or absent), and provided both measurements are based on the same dataset (hence with the same number of independent modes $  ( \kmin /\kmax )^{3}$), the parity-odd signal can be seen only \textit{after} the parity-even signal has been detected with high significance. However, for certain datasets, the noise in the two parity sectors could be quite different, for example in cases where systematic and instrumental noises are not expected to break parity or to do so by a small amount. \\

\paragraph{Small speed of sound} It is interesting to see how the above discussion changes if $\phi$ had a speed of sound $c_s$, and in particular in the case in which $c_s\ll1$. We assume the same interactions as in \eqref{lint} but the following kinetic term
\begin{align}
    \mathcal{L}_2=\frac12 \left[\dot \phi^2-c_s^2 (\partial_i\phi)^2 \right]\,.
\end{align}
To find the parity-odd loop trispectrum and the parity-even tree-level trispectrum we have to decorate \eqref{bpe} and \eqref{bpo} with the appropriate factors of $c_s$. We can proceed by using dimensional analysis, separating energy $E$ (with $E\sim$ time $^{-1}$) from momentum $P$ (with $P\sim$ length$^{-1}$). Noticing that $ \phi(\bfk) \sim E^{-1/2}P^{-3/2}$ we find\footnote{These factors of $c_s$ can also be understood as follows. We get: $c_s^{-12}$ from the normalization of four power spectra $P\sim H^2/(c_s k)^3$; $c_s^{-1}$ for each spatial derivative and no $c_s$ for time derivatives; and $c_s^{3(V-I)}=c_s^{3(1-L)}$ for a diagram with $V$ vertices ($\d\eta/\eta^4\sim c_s^3$), $I$ internal lines $G\sim P \sim c_s^{-3}$ and $L$ loops.}
\begin{align}
\BPE &\sim \left(  \frac{H}{\Lambda_{PE}}\right)^{6}\frac{H^{4}}{k^{9}} \frac{1}{c_s^{13}}\,,\\
\BPO&\sim\left(  \frac{H}{\Lambda_{\text{PO}}}\right)^{9}\left(  \frac{H}{\Lambda_{\text{PE}}}\right)^{6}\frac{H^{4}}{k^{9}} \frac{1}{16\pi^{2}} \frac{1}{c_s^{25}}\,.
\end{align}
For the relative signal-to-noise ratio this gives us
\begin{align}\label{withcs}
\frac{\left(  S/N\right)_{\text{PO}}}{\left(  S/N \right)_{\text{PE}}}=\left(  \frac{H}{\Lambda_{\text{PO}}}\right)^{9}\frac{1}{16\pi^{2}} \frac{1}{c_s^{12}}\,.
\end{align}
This might get your hopes up that a small $c_s$ makes the loop signal in $\BPO$ dominant over the tree-level one in $\BPE$. However, for small $c_s$ the strong coupling scale of the theory is also lowered. Indeed, following \cite{Baumann:2011su} we estimate the strong coupling momentum scale $\ksc$ by demanding that a tree-level 2-to-2 scattering amplitude respects partial wave unitarity. We find $\ksc\sim \Lambda_{\text{PO}}c_s^{1/3}$. Demanding that this cutoff is larger than the wavelength of perturbations at horizon crossing, $\ksc\gg H/c_s$ we find $H/\Lambda_{\text{PO}}\ll c_s^{4/3}$. This can be used in \eqref{withcs} so see that a small $c_s$ has no effect on the upper bound of the relative signal-to-noise ratio, which again has to be much smaller than one. Hence our conclusion that the dominant signal-to-noise ratio is always in the tree-level parity-even trispectrum is unchanged even in the present of a small speed of sound.

In summary, the one-loop signal-to-noise ratio in the parity-odd trispectrum is in general dominated by that in the tree-level parity-even trispectrum, but can be interesting when the two sectors are probed with different accuracy and with different noise and systematic.


\section{Naturalness constraints}\label{sec:natural}

In this section, we discuss constraints from technical naturalness on single-clock inflation with both parity-even and parity-odd interactions. Our main findings are summarized in \eqref{pcsfinal} with our conventions defined in the Lagrangian in \eqref{pcs} \\

An effective field theory (EFT) is said to be technically natural if all of the dimensionless Wilson coefficients are of order unity, except those that enhance the symmetry of the action when taken to zero, which may be small \cite{tHooft:1979rat}. A theory is said to be radiatively stable if counter-terms due to loop diagrams do not spoil the hierarchy between lower- and higher-order operators in the action.
One way to ensure radiative stability in a technically natural EFT is to adopt a suitable power counting scheme, which prescribes which dimensionful parameters should appear alongside the Wilson coefficients of each term in the action. Here we will extend the nice analysis in \cite{Grall:2020tqc} to account for both parity-odd and parity-even interactions in the EFT of single-field inflation.

Before developing a power-counting scheme, let's set some expectations. The parity-odd sector can renormalise the parity-even sector, but not vice versa, because a product of an even number of parity-odd operators is parity-even, but no product of parity-even operators is parity-odd; therefore, we expect to be able to make the parity-odd couplings much weaker than the parity-even ones in a radiatively stable manner, but not vice versa. Let's see how this is borne out in a consistent power counting scheme. \\

To maintain a certain generality we consider the following structure of interactions,
\begin{equation}\label{pcs}
   \mathcal{L} = -\frac{1}{2} (\partial \phi)^2 + \frac{\varepsilon_{PE} \Lambda_{PE}^4}{g_{PE}^2} \mathcal{L}_{PE} \qty(\frac{g_{PE} \phi}{\Lambda_{PE}}, \frac{\partial}{\Lambda_{PE}}) + \frac{\varepsilon_{PO} \Lambda_{PO}^4}{g_{PO}^2} \mathcal{L}_{PO} \qty(\frac{g_{PO} \phi}{\Lambda_{PO}}, \frac{\partial}{\Lambda_{PO}})
\end{equation}
Here $\Lambda_{\text{PE,PO}}$ are the scales suppressing higher derivatives in the parity-even and parity-odd sectors, $\Lambda_{\text{PE}}/g_{\text{PE}}$ and $\Lambda_{\text{PO}}/g_{\text{PO}}$ suppress additional powers of the field and finally $ \e_\text{PE,PO}$ control the overall size of the parity-even and odd sectors. Notice that this is a slightly more sophisticated version of the often employed single-scale power counting scheme where powers of derivative and fields are accompanied by the same scale. The introduction of two different scales via the dimensionless parameters $g_{\text{PE,PO}}$ allows us to account for the fact that loop diagrams can change the number of fields in an interaction but not the number of derivatives (for massless fields in dim reg). The bottom line is that the power counting scheme in \eqref{pcs} is only radiatively stable if 
\begin{align}\label{pcsfinal}
    g_{\text{PO}}& \approx g_{\text{PE}} \ll 4\pi & \Lambda_{\text{PO}} &\approx \Lambda_{\text{PE}} &\varepsilon_{\text{PO}}^2 &\lesssim \varepsilon_{\text{PE}}
\end{align}
Setting the parity-odd sector to zero recovers the power counting scheme discussed in \cite{Grall:2020tqc}. Perturbativity requires $\Lambda_\text{PE,PO}\ll H$. This result is derived as follows.\\

\begin{figure}[h]
   \centering
   \includegraphics[scale=1.25]{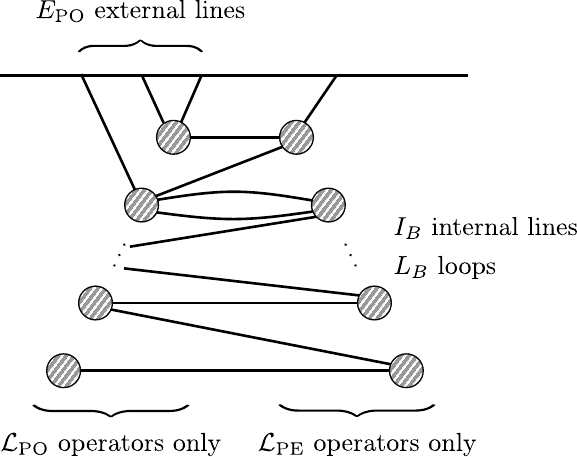}
   \hspace{1.25cm}
   \includegraphics[scale=1.25]{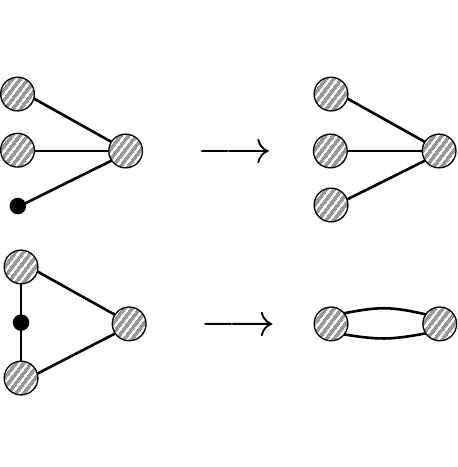}
   \caption{(\textit{Left}) General form of the big diagram.
   Each striped circle represents a maximal connected subdiagram containing operators from only one of $\mathcal{L}_\text{PO}$ or $\mathcal{L}_\text{PE}$ (the little diagrams).
   The big diagram takes the form of a connected bipartite multigraph.
   (\textit{Right}) Illustration of the inductive procedure showing that all connected diagrams are of this form.
   If a new PO operator (the small black circle) is added, it either (\textit{upper right}) forms a new little diagram of its own, if it is not connected directly to other PO little diagrams, or (\textit{lower right}) it merges with any other PO little diagrams it is connected to; either way, the bipartite multigraph structure persists.}
   \label{fig:naturalness diagram}
\end{figure}

\paragraph{Derivation} Consider a general connected Feynman diagram in this theory.
It can be split into connected subdiagrams consisting of vertices from either $\mathcal{L}_{\text{PO}}$ or $\mathcal{L}_{\text{PE}}$ only, together with lines connecting only subdiagrams of the first kind to subdiagrams of the second\footnote{This can be proven inductively: the claim is true for a diagram with one vertex; if a vertex is added to a diagram of this form, it remains of the same form as long as the diagram remains connected.}.
In other words, the diagram, at this level of abstraction (`the big diagram'), looks like a connected bipartite multigraph, where the vertices are in fact connected Feynman diagrams (`the little diagrams').
This is illustrated in Figure \ref{fig:naturalness diagram}.

The strategy will be to write down the general form of the counterterm generated by the big diagram in terms of the properties of the individual little diagrams, then to use the topology of the big diagram to simplify the expression and derive criteria for radiative stability.

If each little diagram $\mathcal{D}_i$ has $N_i$ vertices, $D_i$ derivatives, $L_i$ loops, $F_i$ lines internal to the big diagram (the horizontal lines above) and $E_i$ lines external to the big diagram, the scaling (for a parity-odd (PO) little diagram) is \cite{Grall:2020tqc}
\begin{equation}
   \mathcal{D}_i \sim \frac{\varepsilon_{\text{PO}}^{N_i} \qty(\frac{g_{\text{PO}}}{2\pi})^{2L_i} g_{\text{PO}}^{E_i + F_i - 2} } {\Lambda_{\text{PO}}^{-2(1 + N_i) + E_i + D_i + 2L_i}}.
\end{equation}
There is an analogous expression for PE little diagrams. Now we multiply the scaling of all the little diagrams together. Consider the case in which we there are $N_V^{\text{PO}}$ total PO vertices, $L^{\text{PO}}$ loops, $D^{\text{PO}}$ derivatives, and $E^{\text{PO}}$ external lines on PO vertices, with analogous quantities defined for the PE vertices, and there are $N_B^{\text{PO}}$ PO little diagrams in the big diagram, $N_B^{\text{PE}}$ PE little diagrams, and $L_B$ and $I_B$ loops and internal lines, respectively, in the big diagram. Then the big diagram scales as
\begin{align}
   \mathcal{D}_B &\sim \qty(\frac{1}{4\pi})^{2L_B} (\varepsilon_{\text{PO}})^{N_V^{\text{PO}}} (\varepsilon_{\text{PE}})^{N_V^{\text{PE}}} (g_{\text{PO}})^{E^{\text{PO}} - 2N_V^{\text{PO}}} (g_{\text{PE}})^{E^{\text{PE}} - 2N_V^{\text{PE}}} \nonumber \\
   & \quad \times \qty(\frac{g_{\text{PO}} g_{\text{PE}}}{\Lambda_{\text{PO}} \Lambda_{\text{PE}}})^{I_B} \frac{1}{\Lambda_{\text{PO}}^{A_{\text{PO}}}} \frac{1}{ \Lambda_{\text{PE}}^{A_{\text{PE}}} }, 
\end{align}
where
\begin{align}   
   A_{\text{PO}} &= -2(N_B^{\text{PO}} + N_V^{\text{PO}}) + E^{\text{PO}} + D^{\text{PO}} + 2L^{\text{PO}}, \\
   A_{\text{PE}} &= -2(N_B^{\text{PE}} + N_V^{\text{PE}}) + E^{\text{PE}} + D^{\text{PE}} + 2L^{\text{PE}}.
\end{align}
This diagram gives rise to a counter-term proportional to
\begin{align}
   \mathcal{O}_{CT} &\sim \qty(\frac{1}{4\pi})^{2L_B} (\varepsilon_{\text{PO}})^{N_V^{\text{PO}}} (\varepsilon_{\text{PE}})^{N_V^{\text{PE}}} (g_{\text{PO}})^{E^{\text{PO}} - 2N_V^{\text{PO}}} (g_{\text{PE}})^{E^{\text{PE}} - 2N_V^{\text{PE}}} \nonumber \\
   & \quad\qty(\frac{g_{\text{PO}} g_{\text{PE}}}{\Lambda_{\text{PO}} \Lambda_{\text{PE}}})^{I_B} \frac{\partial^{A_{\text{PO}}}} {\Lambda_{\text{PO}}^{A_{\text{PO}}}} \frac{ \partial^{A_{\text{PE}}}}{ \Lambda_{\text{PE}}^{A_{\text{PE}}} } \partial^{4 + 2 I_B} \phi^{E_B},
\end{align}   
where the Euler identity for the big diagram has been liberally applied, and enough derivatives have been added to give the operator the correct dimensions.
If the operator that is renormalised has parity $Q$, it looks like
\begin{align}
   \mathcal{O} \sim \varepsilon_Q \frac{\Lambda_Q^4}{g_Q^2} \frac{(g_Q)^{E_B}}{\Lambda_Q^{X}} \partial^D \phi^{E_B}\,,
\end{align}
where $X$ is the total number of powers of $\Lambda_{\text{PO}}$ and $\Lambda_{\text{PE}}$ in $\mathcal{O}_{CT}$ and $D$ is the total number of derivatives.
In the case where $\mathcal{O}$ is parity-odd, one finds
\begin{align}
   \label{eq:CT ratio PO}
   \frac{\mathcal{O}_{CT}}{\mathcal{O}} & \sim \qty(\frac{g_{\text{PO}}}{4\pi})^{L_B + 2L^{\text{PO}}} \qty(\frac{g_{\text{PE}}}{4\pi})^{L_B + 2L^{\text{PE}}} (g_{\text{PO}} g_{\text{PE}})^{N_B - 1} (g_{\text{PO}})^{-2N_B^{\text{PO}}} (g_{\text{PE}})^{-2N_B^{\text{PE}}}  \qty(\frac{g_{\text{PE}}}{g_{\text{PO}}})^{E^{\text{PE}}} \nonumber \\
   & \quad   g_{\text{PO}}^2 (\varepsilon_{\text{PO}})^{N_V^{\text{PO}} - 1} (\varepsilon_{\text{PE}})^{N_V^{\text{PE}}}  \qty(\frac{\Lambda_{\text{PO}}}{\Lambda_{\text{PE}}})^{N_B + L_B - 1 - 2N_B^{\text{PE}} + E^{\text{PE}} + D^{\text{PE}} + 2L^{\text{PE}}},
\end{align}
and, further,
\begin{equation} \qty(g_{\text{PO}}g_{\text{PE}})^{N_B - 1} (g_{\text{PO}})^{-2N_B^{\text{PO}}} (g_{\text{PE}})^{-2N_B^{\text{PE}}} = \qty(\frac{g_{\text{PO}}}{g_{\text{PE}}})^{N_B^{\text{PE}}}  \qty(\frac{g_{\text{PE}}}{g_{\text{PO}}})^{N_B^{\text{PO}}} (g_{\text{PO}} g_{\text{PE}})^{-1}.
\end{equation}
Similarly, in the case where $\mathcal{O}$ is parity-even, 
\begin{align}
   \frac{\mathcal{O}_{CT}}{\mathcal{O}} & \sim \qty(\frac{g_{\text{PO}}}{4\pi})^{L_B + 2L^{\text{PO}}} \qty(\frac{g_{\text{PE}}}{4\pi})^{L_B + 2L^{\text{PE}}} \qty(\frac{g_{\text{PO}}}{g_{\text{PE}}})^{N_B^{\text{PE}}}  \qty(\frac{g_{\text{PE}}}{g_{\text{PO}}})^{N_B^{\text{PO}}} (g_{\text{PO}} g_{\text{PE}})^{-1}  \qty(\frac{g_{\text{PO}}}{g_{\text{PE}}})^{E^{\text{PO}}} \nonumber \\
   & \quad   g_{\text{PE}}^2 (\varepsilon_{\text{PO}})^{N_V^{\text{PO}} } (\varepsilon_{\text{PE}})^{N_V^{\text{PE}} - 1}  \qty(\frac{\Lambda_{\text{PE}}}{\Lambda_{\text{PO}}})^{N_B + L_B - 1 - 2N_B^{\text{PO}} + E^{\text{PO}} + D^{\text{PO}} + 2L^{\text{PO}}},
   \label{eq:CT ratio PE}
\end{align}
To derive criteria for radiative stability, note the following:
\begin{itemize}
   \item The ratio \eqref{eq:CT ratio PO} is proportional to $(\Lambda_{\text{PO}}/\Lambda_{\text{PE}})^{D{^{\text{PE}}}}$. Therefore, by putting arbitrarily many derivatives in the parity-even little diagrams, this ratio can be made arbitrarily large and ruin radiative stability unless $\Lambda_{\text{PO}} \lesssim \Lambda_{\text{PE}}$.
      Likewise, \eqref{eq:CT ratio PE} is proportional to $(\Lambda_{\text{PE}}/\Lambda_{\text{PO}})^{D{^{\text{PO}}}}$, implying $\Lambda_{\text{PE}} \lesssim \Lambda_{\text{PO}}$.
      Thus, $\Lambda_{\text{PO}} \approx \Lambda_{\text{PE}}$.
   \item The ratio \eqref{eq:CT ratio PO} is proportional to $(g_{\text{PE}}/g_{\text{PO}})^{E^{\text{PE}}}$. $E^{\text{PE}}$ also affects the exponent of $\Lambda_{\text{PO}}/\Lambda_{\text{PE}}$, but this is unity. The ratio can therefore be made arbitrarily large by adding arbitrarily many external lines on the PE side unless $g_{\text{PE}} \lesssim g_{\text{PO}}$.
      Repeating this argument for the ratio \eqref{eq:CT ratio PE} yields $g_{\text{PO}} \lesssim g_{\text{PE}}$, so that $g_{\text{PO}} \approx g_{\text{PE}}$.
\item The ratio \eqref{eq:CT ratio PO} contains a factor $(\varepsilon_{\text{PO}})^{N_V^{\text{PO}} - 1} (\varepsilon_{\text{PE}})^{N_V^{\text{PE}}}$. For a parity-odd big diagram, $N_V^{\text{PO}} \geq 1$, since there must be at least one (and an odd number of) parity-odd operators in the diagram, and the factor never exceeds unity due to the constraints on $\varepsilon_{\text{PE}}$ and $\varepsilon_{\text{PO}}$ separately.
      However, \eqref{eq:CT ratio PE} is more interesting: it contains a factor $(\varepsilon_{\text{PO}})^{N_V^{\text{PO}}} (\varepsilon_{\text{PE}})^{N_V^{\text{PE}} - 1}$.
      When $N_V^{\text{PE}} = 0$, an even number of PO vertices combine to give a PE diagram, which renormalises a PE operator.
      As each of the $\varepsilon$s is $\lesssim 1$ separately, this factor is $\lesssim \varepsilon_{\text{PO}}^2 / \varepsilon_{\text{PE}}$.
      In order that the radiative corrections do not exceed order unity, since all the other factors in \eqref{eq:CT ratio PE} are of order unity or less, $\varepsilon_{\text{PO}}^2 / \varepsilon_{\text{PE}} \lesssim 1$.
      So $\varepsilon_{\text{PO}} \lesssim \sqrt{\varepsilon_{\text{PE}}}\ll 1$ for radiative stability: the interactions in the parity-odd sector can be a little stronger than in the parity-even sector, but not too strong, since the PO sector still renormalises the PE sector.
      The PO interactions can be arbitrarily weaker than the PE interactions, since the PE sector alone cannot renormalise the PO sector.
\end{itemize}

\paragraph{Broken boosts}
We now consider a theory with broken Lorentz boost symmetry and a finite sound speed\footnote{$c=1$ still; $c_s$ is a dimensionless constant.} $c_s$, where the kinetic term is
\begin{equation}
   -\frac{1}{2} Z^{\mu \nu} \partial_\mu \phi \partial_\mu \phi = \frac{1}{2} \left( (\partial_t \phi)^2 - c_s^2 ( \partial_i \phi )^2 \right)\,,
\end{equation}
and the Lagrangian takes the form
\begin{align}
   \mathcal{L} &= \sqrt{-Z}\left[-\frac{1}{2} Z^{\mu \nu} \partial_\mu \phi \partial_\nu \phi + \frac{\varepsilon_{\text{PE}} \Lambda_{\text{PE}}^4}{g_{\text{PE}}^2} \mathcal{L}_{\text{PE}} \qty(\frac{g_{\text{PE}} \, \phi}{\Lambda_{\text{PE}}}, \frac{\partial}{\Lambda_{\text{PE}}},  \frac{g_{t\,\text{PE}}  \, \partial_t}{\Lambda_{\text{PE}}}, Z^{\mu \nu}) + \right. \\
   & \quad \left. \frac{\varepsilon_{\text{PO}} \Lambda_{\text{PO}}^4}{g_{\text{PO}}^2} \mathcal{L}_{\text{PO}} \qty(\frac{g_{\text{PO}} \, \phi}{\Lambda_{\text{PO}}}, \frac{\partial}{\Lambda_{\text{PO}}},  \frac{g_{t\,\text{PE}} \, \partial_t}{\Lambda_{\text{PE}}}, Z^{\mu \nu}) \right].
\end{align}
Here, $g_{t\,\text{PO}}$ and $g_{t\,\text{PE}}$  parameterise the strength of the boost-breaking, and index contractions are performed with $Z^{\mu \nu}$.
The factor $\sqrt{-Z} = c_s^{-3}$.
In their investigation of EFTs with broken boosts with only one $g_t$ parameter, \cite{Grall:2020tqc} found that $g_t \lesssim 1$.

With parity violations, the effect on the power counting of boost-breaking amounts to multiplying the big diagram by $(g_{t\,\text{PE}})^{T^{\text{PE}}}(g_{t\,\text{PO}})^{T^{\text{PO}}}$, where the $T$s denote the numbers of time derivatives in the little diagrams.
Time derivatives can be converted to spatial derivatives in loops, but not vice versa; if this happens to $C^{\text{PE}}$ time derivatives from the PE vertices and  $C^{\text{PO}}$ time derivatives from the PO vertices, for a PO big diagram, the counterterm ratio \eqref{eq:CT ratio PO} is multiplied by
\begin{equation}
   (g_{t\,\text{PO}})^{C^{\text{PE}} + C^{\text{PO}}} \qty(\frac{g_{t\,\text{PE}}}{g_{t\,\text{PO}}})^{T^{\text{PE}}}.
\end{equation}
Analogously to the previous arguments, for radiative stability, $g_{t\,\text{PE}} \approx g_{t\,\text{PO}}$ and both are $\lesssim 1$.

Additional constraints on the Wilson coefficients were found in \cite{Grall:2020tqc} when the breaking of Lorentz boosts is spontaneous and in the EFT of inflation.
Some of these constraints impose the requirement that the action non-linearly realise Lorentz boosts; others follow by considering radiative stability in this context.
However, these additional constraints do not combine in an interesting way with the presence of a parity-odd sector; no further free parameters are introduced that could differ between the parity-odd and parity-even sectors, and in any case, following the arguments above, we would generally expect any such parameters to be constrained to equality.

\paragraph{Signal-to-noise in the EFT of inflation}
The power-counting scheme nevertheless takes on a slightly different form in the EFT of inflation (EFTI) \cite{Cheung:2007st} in the decoupling limit, as the operators that appear must all be EFTI building blocks (details are omitted here for brevity).
Again extending the work of \cite{Grall:2020tqc}, the Lagrangian takes the form
\begin{align}
   \mathcal{L} &= \sqrt{-Z} \left[ -\frac{1}{2} Z^{\mu \nu} \partial_\mu \pi \partial_\nu \pi + \frac{\Lambda^4}{g_\pi^2} \varepsilon_{\text{PO}} \mathcal{L}_{\text{PO}}\left( \frac{g_\pi^2 c_s^2 f_\pi^4}{ \Lambda^4} \delta g^{00}, \frac{\nabla_\mu}{\Lambda}, g_t n_\mu, Z^{\mu \nu} \right) + \text{ PE interactions} \right],
\end{align}
where $Z^{00} = -1$, $Z^{ij} = c_s^2 \delta^{ij} / a^2$, $\sqrt{-Z} = c_s^{-3}$, $f_\pi^4=2|\dot H|\Mpl^2 c_s$ is the decay constant of the Goldstone boson $\pi$, which is normalised to have mass dimension 1 so that $x^0 \mapsto x^0 + \pi/f_\pi^2$ represents a time diffeomorphism\footnote{This means that every $\pi$ in the interaction vertices appears with a factor of $f_\pi^{-2}$}, $\delta g^{00}$ is a metric fluctuation, and $n_\mu$ is a normal to spacelike hypersurfaces; these quantities can be expressed in terms of derivatives of $\pi$.
In particular, the extrinsic curvature operator ${K^\mu}_\nu = (g^{\mu \sigma}+ n^\mu n^\sigma)  \nabla_\sigma n_\nu $ scales as $g_t^3/\Lambda$, and its fluctuations ${\delta K^\mu}_\nu$ have the same scaling in the power-counting scheme.

The coefficients of the power counting scheme are constrained such that
\begin{equation}
   \label{eq:gt cs ratio}
   \frac{g_t}{c_s} = g_\pi c_s \frac{f_\pi^2}{\Lambda^2}.
\end{equation}
This can be useful in simplifying quotients of signal-to-noise ratios.
The parity-even operator can be written in various ways using EFTI building blocks. One possibility is to consider an operator that starts at quartic order in $\pi$, such as for example $(\delta g^{00})^2(\delta K_{ij})^2$. Another possibility is to consider a combination building blocks that starts at lower order, such as for example $(\delta K_{ij})^2$. In the former case one can invoke naturalness to argue that lower-order operators should give a larger signal-to-noise ratio, for example to the parity-even bispectrum. Instead here we will consider the latter possibility, where a lower-order cubic interaction is dictated by non-linearly realized boosts. Naturalness provides similar bounds, but we don't report them here. We consider the following building blocks to generate the parity-even and parity-odd interactions appearing in our one-loop calculation
\begin{align}\label{PEBB}
   \frac{\varepsilon_\text{PE}}{c_s^3}\frac{\Lambda^4}{g_\pi^2} \left( \frac{g_t^3}{\Lambda} \right)^2 \left( \delta {K^i}_j \right)^2 &\supset \frac{\varepsilon_\text{PE}}{c_s^3}\frac{\Lambda^4}{g_\pi^2} \left( \frac{g_t^3}{\Lambda} \right)^2 \frac{\dot \pi^2 \left( \partial_i \partial_j \pi \right)^2 }{f_\pi^8},\\
   \frac{\varepsilon_\text{PO}}{c_s^3} \frac{\Lambda^4}{g_\pi^2} g_t \varepsilon_{ijk} \left( \frac{g_t^3}{\Lambda} \right)^4 \frac{1}{\Lambda} (\delta K ) (\delta K) \partial (\delta K) (\delta K) &\supset
    \frac{\varepsilon_\text{PO}}{c_s^3} \frac{\Lambda^4}{g_\pi^2} g_t \varepsilon_{ijk} \left( \frac{g_t^3}{\Lambda} \right)^4 \frac{1}{\Lambda} \frac{\partial^2 \pi \partial^2 \pi \partial^3 \pi \partial^2 \pi}{f_\pi^8}\,,
\end{align}
with $\varepsilon_{ijk} = n^\mu \varepsilon_{\mu i j k}$. The index structure in the PO operator is omitted.
The signal-to-noise in the PO trispectrum then scales as\footnote{Note that the $\sqrt{-Z}$ normalisation of the Lagrangian eliminates the factor of $c_s^{3(1 - L)}$ that appeared in correlators in Section \ref{sec:signal}. This normalization also changes the factors of $c_s$ appearing in external propagators, but the factors of $c_s$ due to spatial derivatives remain unaltered.}
\begin{equation}
   \left( \frac{S}{N} \right)_\text{PO} \sim \frac{1}{(2\pi)^9} \left[ \frac{\varepsilon_\text{PO} \varepsilon_\text{PE} g_t^{19}}{(4\pi)^2 g_\pi^4 c_s^{13}} \frac{\Lambda}{H} \left( \frac{H}{f_\pi} \right)^{16} \right] \left( \frac{k_\text{max}}{k_\text{min}} \right)^{3/2}.
\end{equation}
The non-linearly realised boosts, expressed through the building block structure of the EFTI, lead to the following contribution to the bispectrum
\begin{equation}
   \frac{\varepsilon_\text{PE} \Lambda^4}{c_s^3 g_\pi^2} \left( \frac{g_t^3}{\Lambda} \right)^2 \left( \delta {K} \right)^2   \supset  \frac{\varepsilon_\text{PE} \Lambda^4}{c_s^3 g_\pi^2} \left( \frac{g_t^3}{\Lambda} \right)^2 \frac{\dot \pi (\partial^2 \pi)^2 }{f_\pi^6},
\end{equation}
where the exact index structure is unimportant.
The quotient of the signal-to-noise ratios of the PO trispectrum and this contact bispectrum is
\begin{equation}
   \frac{(S/N)_\text{PO}}{(S/N)_{B_3}} \sim \frac{\varepsilon_\text{PO} }{(4\pi)^2 (2\pi)^3} \frac{g_t^{13}}{g_\pi^2 c_s^9} \left( \frac{H}{f_\pi} \right)^{10} 
   = \frac{\varepsilon_\text{PO}}{(4\pi)^2 (2\pi)^3} c_s^{11} g_t^8 g_\pi^3 \left( \frac{H}{c_s \Lambda} \right)^{10}  \lesssim 1,
\end{equation}
where the middle equality follows from \eqref{eq:gt cs ratio}, and $H/(c_s \Lambda) \ll 1$ since $\Lambda$ is the EFT cutoff in momentum space and the EFT description should be valid at the sound horizon. Similar constraints would also emerge from different choices of building blocks in \eqref{PEBB}. In summary, we find that in single-clock inflation, the signal-to-noise in this PO trispectrum is always bounded above by the signal-to-noise in the bispectrum.


\section{Conclusion and outlook}\label{sec:conclusion}

In this work we have pointed out a scenario in which a one loop contribution is the leading term in an observable cosmological correlator. More specifically, we have studied one-loop contributions to the parity-odd sector of scalar correlators from inflation. It is well known that both the scalar power spectrum and bispectrum must be parity even, so a possible violation of parity can only first arise in the scalar trispectrum. This observable has recently attracted some attention \cite{Tong:2022cdz,Cabass:2022rhr,Niu:2022fki,Creque-Sarbinowski:2023wmb} because of related hints from galaxy surveys \cite{Hou:2022wfj,Philcox:2022hkh}. Further investigation however has shown no sign of the signals expected from explicit models \cite{Cabass:2022oap} and bounds from the CMB appear to exclude a primordial origin \cite{Philcox:2022hkh}.

While it will be interesting to further search the data for a parity-odd signal, our interest in the parity-odd trispectrum $\BPO$ stems mainly from theoretical considerations. In \cite{Liu:2019fag,Cabass:2022rhr} it was shown that $\BPO$ must vanish at tree-level for large classes of models. This no-go result assumes unitary time evolution from a Bunch-Davies vacuum and scale invariance, and holds for any number of scalar fields of any mass, or any number of spinning fields with massless or conformally coupled mode functions and parity-even power spectra. This result identifies $\BPO$ as an observable that is \textit{exceptionally sensitive to new physics}, beyond the vanilla models of inflation. Like for any other no-go theorem, the interest lies in building yes-go examples that violate one or more of the assumptions. A variety of models have been proposed that generate a non-vanishing $\BPO$ \cite{Tong:2022cdz,Cabass:2022rhr,Niu:2022fki,Creque-Sarbinowski:2023wmb}. Here we relax the assumption of working at tree-level. We were able to compute a variety of possible parity-odd trispectra, for example in multifield inflation, \eqref{finalss} and \eqref{finalss2}. The most minimal setup we consider is that of single-clock inflation, as described by the EFT of inflation, with a Bunch-Davies vacuum and scale invariance. In this case the final shape was given in \eqref{finalBPO} with relevant definitions given in the rest of Section \ref{sec:single}.\\

There are several interesting directions for future investigation:
\begin{itemize}
    \item Detailed cancellations from the wavefunction to correlation functions. For the one-vertex one-loop diagram of massless scalars in Minkowski space, we have shown that the wavaefunction $\psi_4^{(1L)}$ possesses a $\log k_T$ term. However this term is cancelled by a contribution from $\psi_6^{\text{tree}}$ when we compute the correlator, and the result is analytic in the external kinematics. We have also shown that a similar cancellation occurs for massive scalars in Minkowski space as well. One may ask if such a cancellation occurs when we consider more complicated loop diagrams as well. The analytic structure of the wavefunction for Minkowski space has been studied extensively in \cite{Salcedo2022}, but our results for the one-vertex one-loop calculation seems to suggest that not all singularities in the wavefunction appear in an in-in correlator. It is important to investigate whether these are examples of a more general phenomenon and if yes, to develop a systematic understanding of such cancellations. This issue has a vague resemblance to the Kinoshita Lee Nauenberg (KLN) theorem \cite{Lee:1964is, Kinoshita:1962ur} that determines the faith of IR divergences when computing sufficiently inclusive observables.
    \item On general grounds, one expects that the one-loop trispectrum we computed here represents a genuine quantum effect, as opposed to a classical one. This is follows from the observation that in calculating wavefunction coefficients one finds a factor of\footnote{Every vertex has a factor of $\hbar^{-1}$ and every bulk-bulk propagator a factor of $\hbar$.} $\hbar^{L-1}$ for a diagram with $L$ loops. Things are slightly more subtle for correlators, which correspond to sums of both loop and tree-level wavefunction coefficients. 
    Since tree-level wavefunction coefficients are obtained directly from the classical on-shell action, this seems to suggest that in-in correlators from loop diagrams might contain both classical and quantum contributions. It would be useful to clarify this interpretation. This is related to the first point discussed about where a cancellation takes place between a tree-level and a one-loop wavefunction coefficient, which contribute at the same order in $\hbar$ because of appropriate factors of the power spectrum.
    \item We find it surprising that the parity-odd one-loop trispectra we computed have such a simple structure: they are all just rational functions of the kinematics with only total energy poles. Even for a diagram with two interaction vertices, no partial energy singularities appear in the final result. Moreover, we don't find any branch points, only poles. This would not be the case in the parity-even sector. Moreover, the one-loop $\BPO$ we find could be attributed to a \textit{local} contact interaction with an imaginary coupling. It would be nice to understand the deeper origin of this simplicity, perhaps in the context of open quantum systems. 
    \item Another option to understand the simplicity of $\BPO$ is to compute it using the wavefunction. Since we found that the only contribution comes from interactions on the same side of the in-in contour (either $++$ or $--$), see discussion around \eqref{noB4B}, we expect that $\BPO$ is captured by the one-loop $\psi_4$, with other tree-level contributions cancelling out. Then one can use the cosmological optical theorem at one-loop order \cite{sCOTt} to relate this term to the integral of a tree-level exchange diagram for $\psi_6$. This might shed some light on why the result is simply a rational function\footnote{We are thankful to Sadra Jazayeri for suggesting this approach.}.
    \item It would be interesting to study parity violations in the tensor sector that are induced by additional, potentially massive fields. This could be achieved via direct calculation or using de Sitter \cite{Arkani-Hamed:2018kmz,Baumann:2019oyu,Baumann:2020dch} or boostless \cite{BBBB,Jazayeri:2021fvk,Bonifacio:2022vwa} bootstrap techniques.
\end{itemize}


\section*{Acknowledgements} We would like to thank Giovanni Cabass, Sadra Jazayeri, Sébastien Renaux-Petel and David Stefanyszyn for useful discussions. E.P. has been supported in part by the research program VIDI with Project No. 680-47-535, which is (partly) financed by the Netherlands Organisation for Scientific Research (NWO). M.H.G.L. is supported by the Croucher Cambridge International Scholarship. C.M.\ is supported by Science and Technology Facilities Council (STFC) training grant ST/W507350/1. This work has been partially supported by STFC consolidated grant ST/T000694/1.


\appendix

\section{Feynman rules}\label{app:A}

To compute cosmological correlators, we use in-in Feynman diagrams.
These are discussed in detail in the context of cosmology in, for example, \cite{Chen:2017ryl}.
The Feynman rules to compute $\ev{\phi(\vk_1)\cdots\phi(\vk_m)}$ at some final time $\eta_f$ are
\begin{itemize}
   \item Draw vertices corresponding to the interaction terms in the action.
   \item Label each vertex as time ordered or anti-time ordered.
      Time-ordered vertices are drawn as shaded circles, and anti-time ordered vertices as open circles.
      Also assign to each vertex a conformal time $\eta_a$ with $a=1,\dots,I$ labelling the $I$ internal lines. Diagrams with opposite labelling are related by complex conjugation as in \eqref{DDbar}. 
   \item In addition to factors of $i$ arising from spatial derivatives, each time-ordered vertex carries another factor of $-i$ and each anti-time ordered vertex a factor of $i$.
      Since all possible time orderings are summed over, if a diagram contains an even number of spatial derivatives in total, only its real part appears in the final correlator.
      Similarly, with an odd number of spatial derivatives, only the imaginary part appears.
   \item Draw internal lines connecting the vertices pairwise.
      Assign each internal line a 3-momentum $\bfp$. For each $\phi(\vk_i)$ in the correlator, draw one external line, connecting a vertex to a horizontal line representing the asymptotic future.
      Label each external line with its associated 3-momentum $\vk_i$.
      Write down a 3-momentum conserving $\delta$ distribution for each vertex.
      The total number of lines meeting at each vertex is fixed as usual by the fields appearing in the relevant interaction.
   \item With $f_k(\eta)$ the mode function for a given field, for each line, write down a propagator as follows:
      \begin{align}
         \includegraphics[scale=1.3]{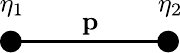} \quad &= G_{++}(\bfp, \eta_1, \eta_2) = f_p(\eta_1) f_p^*(\eta_2) \theta(\eta_1 - \eta_2) + f_p^*(\eta_1) f_p(\eta_2) \theta(\eta_2 - \eta_1)\\
         \includegraphics[scale=1.3]{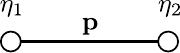} \quad &= G_{--}(\bfp, \eta_1, \eta_2) = f_p^*(\eta_1) f_p(\eta_2) \theta(\eta_1 - \eta_2) + f_p(\eta_1) f_p^*(\eta_2) \theta(\eta_2 - \eta_1)\\
         \includegraphics[scale=1.3]{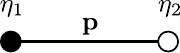} \quad &= G_{+-}(\bfp, \eta_1, \eta_2) = f_p^*(\eta_1) f_p(\eta_2) \\
         \includegraphics[scale=1.3]{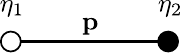} \quad &= G_{-+}(\bfp, \eta_1, \eta_2) = f_p(\eta_1) f_p^*(\eta_2) \\
         \includegraphics[scale=1.3]{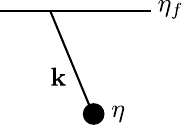} \quad &= G_{+}(\vk, \eta) = f_k^*(\eta) f_k(\eta_f)\\
         \includegraphics[scale=1.3]{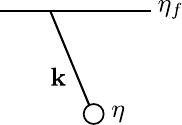} \quad &= G_{-}(\vk, \eta) = f_k(\eta) f_k^*(\eta_f)
      \end{align}
   \item Integrate over all momenta of internal lines and over the conformal times $\eta_i$ of the vertices, with a factor of $a^{1 + d}(\eta_i)$, where $a$ is the scale factor and $d$ the number of spatial dimensions, for each conformal time $\eta_i$.
   \item Multiply by a combinatorial factor depending on the number of ways of contracting fields in the vertices to produce the same diagram.
\end{itemize}


\section{Tensor structures}\label{app:B}

The $I$ tensors that result from the single-field loop integral are totally symmetric and can only depend on the vector $\vb{s}$, so they take the form of sums of products of $s_i$ and $\delta_{ij}$.
The momentum integrals are computed in dimensional regularisation; when the integrand is not a scalar, this makes it more difficult to compute the integral. In this appendix, we determined the coefficients of the products of $s_i$ and $\delta_{ij}$ in terms of contractions of the $I$ tensors with other powers of $s_i$ and $\delta_{ij}$.\\

Since the number of contractions will grow rapidly with the number of indices, it is useful to introduce a diagrammatic way to represent them. The diagrammatic rules are summarised below. Diagrams will be drawn in two columns, with tensors on the \textit{left} representing those contracted into the factors in $I$ on the \textit{right}.
\begin{center}
   \includegraphics[scale=1.35]{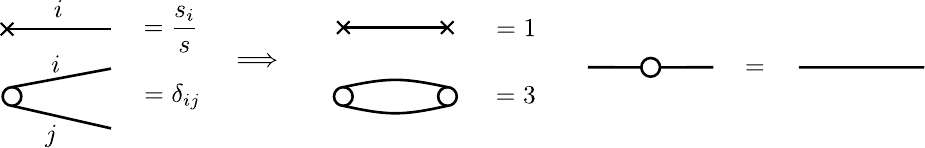}
\end{center}
Since $I$ is totally symmetric, the symmetry factor for each diagram corresponds to the number of distinct ways of assigning labels (indices) to the internal lines whilst preserving the index structure on the left---i.e.\ the indices that meet at any vertex there.
For example, the following diagram, which appears in $\frac{1}{s}s_i \delta_{jk} I_{ijk}$, has symmetry factor 2:
\begin{center}
   \includegraphics[scale=1.35]{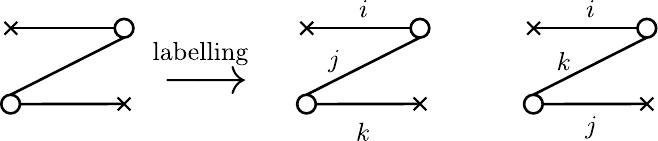}
\end{center}

\paragraph{Three indices}

With three indices, $I$ takes the form
\begin{equation}
   I_{ijk} = I^{(3)}_1 \left( \frac{s_i}{s} \delta_{jk} + \frac{s_j}{s}\delta_{ik} + \frac{s_k}{s} \delta_{ij} \right) + I^{(3)}_3 \frac{s_i s_j s_k}{s^3}.
\end{equation}
In this case, it is straightforward to calculate the relevant traces directly; diagrams are shown below:
\begin{center}
   \includegraphics[scale=1.35]{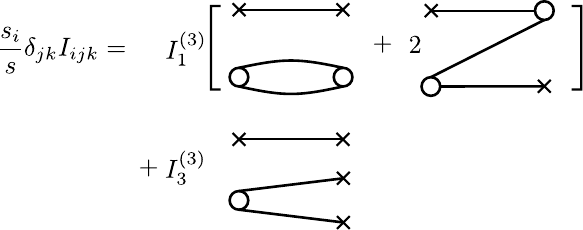}\\
   \vspace{3mm}
   \includegraphics[scale=1.35]{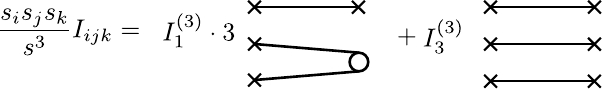}
\end{center}
The result is that
\begin{align}
   T_1&:= \frac{s_i}{s}\delta_{jk}I_{ijk} = 5 I^{(3)}_1 + I^{(3)}_3\\
   T_3&:= \frac{s_i s_j s_k}{s^3} I_{ijk} = 3 I^{(3)}_1 + I^{(3)}_3,
\end{align}
so that
\begin{align}
   I^{(3)}_1 &= \frac{T_1 - T_3}{2} \,, &    I^{(3)}_3 &= \frac{-3T_1 + 5T_3}{2}.
\end{align}

\paragraph{Four indices}

The diagrammatic method simplifies calculation of the $I$ tensors with more indices.
The diagrams themselves are omitted here for brevity.

With four indices,
\begin{equation}
   I_{ijkl} = I^{(4)}_0 \left( \delta_{ij}\delta_{kl} + \text{2 perms} \right) + I^{(4)}_2 \left( \frac{s_i s_j}{s^2} \delta_{kl} + \text{5 perms} \right) + I^{(4)}_4 \frac{s_i s_j s_k s_l}{s^4}.
\end{equation}
Using $T_n$ to denote a trace with $n$ $\frac{s_i}{s}$s, then,
\begin{align}
   T_0 &= \delta_{ij}\delta_{kl} I_{ijkl} = 15 I^{(4)}_0 + 10 I^{(4)}_2 + I^{(4)}_4\,,\\
   T_2 &= \frac{s_i s_j}{s^2} \delta_{kl} I_{ijkl} = 5 I^{(4)}_0 + 8 I^{(4)}_2 + I^{(4)}_4\,, \\
   T_4 &= \frac{s_i s_j s_k s_l}{s^4} I_{ijkl} = 3 I^{(4)}_0 + 6 I^{(4)}_2 + I^{(4)}_4,
\end{align}
so that
\begin{align}
   I^{(4)}_0 &= \frac{1}{8}\left( T_0 -2T_2 + T_4 \right)\,,\\
   I^{(4)}_2 &= \frac{1}{8}\left( -T_0 + 6T_2 - 5T_4 \right)\,,\\
   I^{(4)}_4 &= \frac{1}{8}\left( 3T_0 - 30T_2 + 35T_4 \right).
\end{align}

\paragraph{Five indices}
At this point, the number of terms starts to become large, and the diagrammatic approach begins to pay off.
\begin{equation}
   I_{ijklm} = \frac{I^{(5)}_1}{s} \left( s_i \delta_{jk}\delta_{lm} + \text{14 perms} \right) + \frac{I^{(5)}_3}{s^3}\left( s_i s_j s_k \delta_{lm} + \text{9 perms} \right) + \frac{I^{(5)}_5}{s^5} s_i s_j s_k s_l s_m.
\end{equation}
Using the notation established above, 
\begin{align}
   T_1 &= 35 I^{(5)}_1 + 14 I^{(5)}_3 + I^{(5)}_5\,,\\
   T_3 &= 21 I^{(5)}_1 + 12 I^{(5)}_3 + I^{(5)}_5\,,\\
   T_5 &= 15 I^{(5)}_1 + 10 I^{(5)}_3 + I^{(5)}_5.
\end{align}
This yields
\begin{align}
   I^{(5)}_1 &= \frac{1}{8}\left( T_1 -2T_3 + T_5 \right)\,,\\
   I^{(5)}_3 &= \frac{1}{8}\left( -3T_1 + 10 T_3 - 7T_5 \right)\,,\\
   I^{(5)}_5 &= \frac{1}{8}\left( 15T_1 - 70T_3 + 63T_5 \right).
\end{align}

\paragraph{Six indices}
Now,
\begin{align}
   I_{ijklmn} &= I^{(6)}_0 \left( \delta_{ij} \delta_{kl} \delta_{mn} + \text{14 perms} \right) + \frac{I^{(6)}_2}{s^2} \left( s_i s_j \delta_{kl} \delta_{mn} + \text{44 perms} \right) + \nonumber \\
   & \quad \frac{I^{(6)}_4}{s^4} \left( s_i s_j s_k s_l \delta_{mn} + \text{14 perms} \right) + \frac{I^{(6)}_6}{s^6} s_i s_j s_k s_l s_m.
\end{align}
The required traces are
\begin{align}
   T_0 &= 105 I^{(6)}_0 + 105 I^{(6)}_2 + 21 I^{(6)}_4 + I^{(6)}_6,\\
   T_2 &= 32 I^{(6)}_0 + 77 I^{(6)}_2 + 19 I^{(6)}_4 + I^{(6)}_6,\\
   T_4 &= 21 I^{(6)}_0 + 57 I^{(6)}_2 + 17 I^{(6)}_4 + I^{(6)}_6,\\
   T_6 &= 15 I^{(6)}_0 + 45 I^{(6)}_2+ 15 I^{(6)}_4 + I^{(6)}_6.
\end{align}
Thus,
\begin{align}
   I^{(6)}_0 &= \frac{1}{57}\left( T_0 - 3T_2 + 3 T_4 - T_6 \right),\\
   I^{(6)}_2 &= \frac{1}{456} \left( -5 T_0 + 72 T_2 - 129 T_4 + 62 T_6 \right),\\
   I^{(6)}_4 &= \frac{1}{76} \left( T_0 - 60 T_2 + 155 T_4 - 96 T_6\right), \qq{and}\\
   I^{(6)}_6 &= \frac{1}{152} \left( 5 T_0 + 840 T_2 - 2835 T_4 + 2142 T_6 \right).
\end{align}

\section{Loop computation with cutoff regularization}\label{app:c}
Let us compute the integral \eqref{toyB4A} with cutoff regularization. This integral in cutoff regularization reads:
\begin{multline}
	B_{4A}=2i\text{ Im}\int \frac{d\eta_1}{(-H\eta_1)^{4}}\int \frac{d\eta_2}{(-H\eta_2)^{4}}\int^{\Lambda a(\eta_1)} \frac{d^{3}p}{(2\pi)^3}(i^3(-H\eta_2)^3)(\bfk_1\times\bfk_2\cdot\bfp_1)\\\times(i^2(-H\eta_2)^2(\bfk_3\cdot\bfp_1))G_+(k_1,\eta_1)G_+(k_2,\eta_1)G_{++}(p_1,\eta_1,\eta_2)\\\times G_{++}(p_2,\eta_1,\eta_2)G_+(k_3,\eta_2)G_+(k_4,\eta_2).
\end{multline}
Before proceeding, let us note the following:
\begin{itemize}
    \item The cutoff is $\Lambda a(\eta_2)$ instead of just $\Lambda$. This is because we would like to impose a cutoff on the physical momentum \cite{Senatore:2009cf}, which is the relevant physical quantity, rather than the comoving momentum.
    \item When computing the nested time integrals, the domain for the time integrals also needs to be modified as:
    \begin{equation}
        \int_{-\infty}^{0} d\eta_1 \int_{-\infty}^{\eta_1(1+\frac{H}{\Lambda})}d\eta_2.
    \end{equation}
    This is because if we allow $\eta_2$ to be arbitrarily close to $\eta_1$, we are probing regions that have energy above the cutoff.
\end{itemize}
With this in mind, let us compute the integrals. Doing the angular part of the loop momentum integral give us:
\begin{multline}
    B_{4A}=2i\text{ Im}\frac{(H\eta_0)^4H^5}{64k_1k_2k_3k_4}(\bfk_1\times\bfk_2)\cdot\bfk_3(\partial_{\omega_L})^3(\partial_{\omega_R})^2\\\times\frac{1}{8\pi^2}\frac{1}{6}    \int_{-\infty}^{0}d\eta_1\int_{-\infty}^{\eta_1(1+\frac{H}{\Lambda})}d\eta_2\int_{s}^{\frac{\Lambda}{H\eta_2}}dp_+(p_+^2-s^2)e^{i\omega_L\eta_1}e^{i\omega_R\eta_2}e^{i p_+(\eta_2-\eta_1)}+(\omega_L\leftrightarrow\omega_R).
\end{multline}
Let us define the following:
\begin{equation}
    B_{4A}:=2i\text{ Im}\frac{(H\eta_0)^4H^5}{64k_1k_2k_3k_4}(\bfk_1\times\bfk_2)\cdot\bfk_3(\partial_{\omega_L})^3(\partial_{\omega_R})^2\frac{1}{8\pi^2}\frac{1}{6}A^{\Lambda}_2(\omega_L,\omega_R).\label{midB4}
\end{equation}
Now we compute $A^{\Lambda}_2$. First we do the momentum integral and obtain:
\begin{multline}
    A^{\Lambda}_2=\int_{-\infty}^{0}d\eta_1\int_{-\infty}^{\eta_1(1+\frac{H}{\Lambda})}d\eta_2\frac{1}{(\eta_1-\eta_2)^3}e^{i\omega_L\eta_1}e^{i\omega_R\eta_2}\\
    \left[\frac{i}{H^2\eta_2^2}e^{i\frac{\Lambda(\eta_2-\eta_1)}{H\eta_2}}(-\Lambda^2(\eta_1\eta_2)^2+2iH\eta_2\Lambda (\eta_1-\eta_2)+H^2\eta_2^2(2+s^2(\eta_1-\eta_2)^2)\right.\\
    \left.+e^{is(\eta_2-\eta_1)}(2i+2s(\eta_2-\eta_1))\right]+(\omega_L\leftrightarrow\omega_R).
\end{multline}
The second line is a rapidly oscillating integral as $\Lambda\rightarrow\infty$, and so it averages to zero. Therefore we just need to integrate the third line. Doing the $\eta_2$ integral gives:
\begin{equation}
    A^{\Lambda}_2=\int_{-\infty}^{0}d\eta_1 e^{i\omega_T\eta_1}\left[\frac{-i\Lambda^2}{H^2\eta_1^2}+\frac{\Lambda(s-\omega_R)}{H\eta_1}-i(\omega_R^2-s^2)\text{Ei}\left(\frac{iH\eta_1}{\Lambda}(s+\omega_R)\right)\right]+(\omega_L\leftrightarrow\omega_R).
\end{equation}
We only want the imaginary part of this integral. The first two terms can easily be seen to be real by performing a Wick rotation in $\eta_1$. Notice that the $(\partial_{\omega_L})^3$ operator in \eqref{midB4} bring down factors of $\eta_1$, which ensures convergence of the integral. In addition, we know that the exponential integral has an expansion:
\begin{equation}
    \text{Ei}(z)=\gamma+\log(z)+\sum_{k=1}^{\infty}\frac{z^k}{k\,k!},
\end{equation}
where $\gamma$ is the Euler-Mascheroni constant. The exponential integral $\text{Ei}(z)$ is the sum of a logarithm and an entire function, so we can integrate the series expansion term by term. In particular, we find that since $z=\frac{iH\eta_1}{\Lambda}(s+\omega_R)$ here, the terms in the series are seen to be purely real upon performing a Wick rotation. Keeping the imaginary part leaves us with:
\begin{equation}
    \text{ Im}A^{\Lambda}_2=\int_{-\infty}^{0}d\eta_1 e^{i\omega_T\eta_1}\pi(s^2-\omega_R^2)+(\omega_L\leftrightarrow\omega_R)=\frac{\pi}{\omega_T} (\omega_R^2-s^2+\omega_L^2-s^2).\label{A2lambda}
\end{equation}
Note again that $(\partial_{\omega_L})^3$ ensures convergence of the $\log \eta$ term, which hence does not contribute to the imaginary part. Notice that the final result is independent of the cutoff $\Lambda$. Substituting \eqref{A2lambda} into \eqref{midB4} and doing the $\partial_{\omega_L}$ and $\partial_{\omega_R}$ derivatives returns \eqref{toy}, which is the result from dim reg.
\bibliographystyle{JHEP}
\bibliography{refs}
\end{document}